\documentclass[11pt,twoside,english]{extarticle}
\usepackage{ae,aecompl}
\usepackage[T1]{fontenc}
\usepackage[a4paper]{geometry}
\usepackage{color}
\usepackage[swedish,finnish,english]{babel}
\usepackage{array}
\usepackage{textcomp}
\usepackage{amsmath}
\usepackage{amssymb}
\usepackage{graphicx}
\usepackage[authoryear]{natbib}
\usepackage[unicode=true,pdfusetitle,
 bookmarks=true,bookmarksnumbered=false,bookmarksopen=false,
 breaklinks=false,pdfborder={0 0 0},backref=false,colorlinks=false]
 {hyperref}

\makeatletter

\newcommand{\lyxmathsym}[1]{\ifmmode\begingroup\def\b@ld{bold}
  \text{\ifx\math@version\b@ld\bfseries\fi#1}\endgroup\else#1\fi}

\providecommand{\tabularnewline}{\\}

\date{}

\usepackage[nolist,nohyperlinks]{acronym}

\usepackage{eurosym}
\@ifundefined{rotatebox}
{\usepackage{graphicx}}{}
\definecolor{green}{RGB}{59, 211, 39}
\definecolor{red}{RGB}{192, 0, 0}
\definecolor{blue}{RGB}{0, 176, 240}
\definecolor{cyan}{RGB}{255, 0, 0}
\definecolor{magenta}{RGB}{0, 0, 255}
\definecolor{gray}{RGB}{211, 211, 211}
\definecolor{yellow}{RGB}{128, 128, 128}
\newenvironment{widePar}[2]%
 {\begin{list}{}{\leftmargin#1\rightmargin#2}\item{}}{\end{list}}
\usepackage[margin=11pt,font=small,labelfont=bf,labelsep=colon]{caption}
\usepackage{enumitem}
\setlist{itemsep=0pt}

\definecolor{CB_red}{RGB}{228, 26, 28}
\definecolor{CB_blue}{RGB}{55, 126, 184}
\definecolor{CB_green}{RGB}{77, 175, 74}
\definecolor{CB_purple}{RGB}{152, 78, 163}
\definecolor{CB_orange}{RGB}{255, 127, 0}
\definecolor{CB_yellow}{RGB}{255, 255, 51}
\definecolor{CB_brown}{RGB}{166, 86, 40}
\definecolor{CB_pink}{RGB}{247, 129, 191}

\long\def\symbolfootnote[#1]#2{\begingroup%
\def\thefootnote{\fnsymbol{footnote}}\footnote[#1]{#2}\endgroup}

\makeatother

\begin{document}
\selectlanguage{english}

\title{\noindent Macroprudential oversight, risk communication and visualization%
\thanks{This research has been supported by a grant from the SWIFT Institute.
The author wants to thank Mark Flood, Robert Hofmeister, John Kronberg,
Victoria L. Lemieux, Samuel Rönnqvist, Mikael Sand, Peter Ware and
members of the SWIFT Institute\textquoteright{}s Advisory Council
for insightful comments and discussions. The VisRisk platform for
interactive and analytical applications has been produced in conjunction
with and is property of infolytika, and can be found here: \protect\href{http://vis.risklab.fi/}{http://vis.risklab.fi/}.
VisRisk is open to submissions of systemic risk data, indicators and
models for visualization, which should be directed to the author of
this paper. The paper has also benefited from feedback from conference
and seminar participants at the Data Mining and Knowledge Management
Laboratory, Åbo Akademi University on August 27, 2013 in Turku, Finland,
the {\small $11^{\text{th}}$} Payment and Settlement System Simulation
Seminar on 29--30 August, 2013 at the Bank of Finland, Helsinki, Universitat
Pompeu Fabra on 18 September, 2013 in Barcelona, the Deutsche Bundesbank
on 28 October, 2013 in Frankfurt am Main, the {\small $7^{\text{th}}$}
International Conference on Computational Financial Econometrics on
14 December, 2013 at UCL and LSE in London, Arcada University of Applied
Sciences on November 14, 2013 in Helsinki, Eindhoven University of
Technology on January 7, 2014, the Future of Financial Standards conference
by SWIFT Institute, SWIFT's Standards Forum and LSE on March 25, 2014
in London, the Center of Excellence SAFE, Goethe University on 7 May,
2014 in Frankfurt, the Systemic Risk Center at LSE on May 16, 2014
in London, and at the ECB Financial Stability seminar on 25 June,
2014 in Frankfurt.%
}}

\maketitle
\noindent \begin{center}
{\Large Peter Sarlin}\symbolfootnote[2]{Center of Excellence SAFE at Goethe University Frankfurt and RiskLab Finland at IAMSR, Åbo Akademi University and Arcada University of Applied Sciences. Correspondence to: Peter Sarlin, Goethe University, SAFE, Grüneburgplatz 1, 60323 Frankfurt am Main, Germany. E-mail: psarlin@abo.fi.}
\par\end{center}

\vspace{-0.5cm}

\noindent \begin{center}
Goethe University Frankfurt \\
RiskLab Finland
\par\end{center}

\vspace{0.7cm}

\textbf{\hspace{-0.8cm}}\emph{Abstract}: This paper discusses the
role of risk communication in macroprudential oversight and of visualization
in risk communication. Beyond the soar in data availability and precision,
the transition from firm-centric to system-wide supervision imposes
vast data needs. Moreover, except for internal communication as in
any organization, broad and effective external communication of timely
information related to systemic risks is a key mandate of macroprudential
supervisors, further stressing the importance of simple representations
of complex data. This paper focuses on the background and theory of
information visualization and visual analytics, as well as techniques
within these fields, as potential means for risk communication. We
define the task of visualization in risk communication, discuss the
structure of macroprudential data, and review visualization techniques
applied to systemic risk. We conclude that two essential, yet rare,
features for supporting the analysis of big data and communication
of risks are analytical visualizations and interactive interfaces.
For visualizing the so-called macroprudential data cube, we provide
the VisRisk platform with three modules: \emph{plots}, \emph{maps}
and \emph{networks}. While VisRisk is herein illustrated with five
web-based interactive visualizations of systemic risk indicators and
models, the platform enables and is open to the visualization of any
data from the macroprudential data cube.\\
\textbf{}\\
\textbf{\hspace{-0.8cm}}\emph{Keywords}: Macroprudential oversight,
risk communication, visualization, analytical visualization, interactive
visualization, VisRisk

\thispagestyle{empty}

\newpage{}

\section*{Non-technical summary}

The policy objective of safeguarding financial stability, which is
addressed through macroprudential oversight of the financial system,
is currently being accepted and implemented within governmental authorities
and supervisors. Beyond the soar in availability and precision of
data, the transition from firm-centric to system-wide supervision
imposes obvious data needs when analyzing a large number of entities
and their constituents as a whole. As central tasks ought to be timely
and accurate measurement of systemic risks, big data and analytical
models and tools become a necessity. While analytics might aid in
automated modeling, one approach to dealing with complex data and
modeling problems is to improve end users' understanding of them in
order to tap into their expertise. This points towards means that
support \textit{\emph{disciplined and structured judgmental analysis}}
based upon policymakers' experience and domain intelligence. Further,
the mandates of macroprudential supervisors have to date been stressing
or even limited to communication, issuing warnings and giving recommendations,
which boils down to an emphasis on broad and effective communication
of timely information related to systemic risks.

Systemic risk has commonly been distinguished into three categories:
(\emph{i}) build-up of widespread imbalances, (\emph{ii}) exogenous
aggregate shocks, and (\emph{iii}) spillover and contagion. With the
aim of mitigating system-wide risks, macroprudential oversight is
commonly comprised into a process, where key tasks include (\emph{i})
risk identification, (\emph{ii}) risk assessment, and (\emph{iii})
policy assessment, implementation and follow-up. As a soft policy
intervention, risk communication concerns the overall task of spreading
broadly and effectively timely information related to systemic risks,
as well as other vulnerabilities concerning the financial system and
its macro-financial environment. Fortunately, policymakers and regulators
have access to a broad toolbox of analytical models to measure and
analyze system-wide threats to financial stability. The tasks of these
tools can be mapped to the above listed three forms of systemic risk:
(\emph{i}) early-warning models and indicators, (\emph{ii}) macro
stress-test models, and (\emph{iii}) contagion and spillover models.
While the first aids in risk identification, the second and third
approaches provide means for risk assessment. Yet, this points out
a mismatch between the current objectives and needs and the available
tools: while a key task is the communication of risks, the toolbox
of analytical models lacks a focus on approaches that support human
understanding.

The term visualization has a wide meaning and relates to a number
of interdisciplinary topics, in particular information visualization
and visual analytics. The rationale behind the use of visual representations
and their usefulness relates to traits of the human visual system.
Visualization can be seen as a type of cognitive support or amplification,
which leads to a focus on strengths and weaknesses of human perception.
This highlights the importance of principles for designing visuals
that meet the demands of the human visual system. Next, the utilized
techniques for visualization can be divided into two types: graphical
representations of data and means for interaction. While the former
can be summarized in various categories of visualization techniques,
such as per output and data, the latter refer to how the user can
interact with or manipulate the displayed data, such as zooming or
panning, which often has its basis in one or more graphical displays
for enabling more fl{}exibility to explore data. This invokes two
questions: \emph{1. (2. how) would tapping into visualization support
risk communication in macroprudential oversight?}

This paper discusses the role of visualization in macroprudential
oversight at large, especially for the purpose of risk communication.
Risk communication comprises two tasks. Internal communication concerns
spreading information about systemic risks within but at various levels
of the organization, such as among divisions, groups or analysts,
whereas external communication refers to the task of disseminating
information about systemic risks to the general public. In this paper,
we mainly focus on the background and theory of information visualization
and visual analytics, as well as techniques provided within these
disciplines, as potential means for risk communication. The topic
of visualization is in this paper discussed from three viewpoints:
(\emph{i}) we define the task of visualization in risk communication,
(\emph{ii}) present a so-called macroprudential data cube and discuss
its structure, and (\emph{iii}) review visualization techniques applied
to systemic risk. This provides an overview of which tasks should
be supported by visualization and the underlying data to be visualized.
Eventually, the discussion boils down to two essential, but to date
rare, features for supporting the analysis of big financial data and
the communication of risks: analytical visualizations and interactive
interfaces.

For visualizing the macroprudential data cube through analytical and
interactive visualization, we provide the VisRisk platform with three
modules: \emph{plots}, \emph{maps} and \emph{networks}.%
\footnote{The VisRisk platform for interactive and analytical applications can
be found here:

\href{http://vis.risklab.fi/}{http://vis.risklab.fi/}%
} \emph{Plots} focus on interactive interfaces for representing large
amounts of data. While \emph{maps} provide analytical means for representing
the three standard dimensions of a data cube in simple formats, \emph{networks}
aim at visualization of the fourth data cube dimension of interlinkages.
While VisRisk enables and is open to the visualization of any data
from a macroprudential data cube, the platform is herein illustrated
with five web-based interactive visualizations of systemic risk indicators
and models, of which three make use of analytical visualizations.
First, we make use of analytical techniques for data and dimension
reduction to explore high-dimensional systemic risk indicators and
time-varying networks of linkages. Second, this paper adds interactivity
to not only dashboards of standard risk indicators and early-warning
models, but also to the analytical applications. The ultimate aim
of VisRisk, and this paper at large, is to provide a basis for the
use of visualization techniques, especially those including analytical
and interactive features, in macroprudential oversight in general
and risk communication in particular.

\newpage{}

\begin{widePar}{+2.0in}{-0.0in}

\textit{\footnotesize \textquotedblleft{}}\emph{\footnotesize In the
absence of clear guidance from existing analytical frameworks, policy-makers
had to place particular reliance on our experience. Judgement and
experience inevitably played a key role. {[}...{]}}\textit{\emph{\footnotesize{}
}}\textit{\footnotesize But relying on judgement inevitably involves
risks. We need macroeconomic and financial models to discipline and
structure our judgemental analysis. How should such models evolve?\textquotedblright{}}{\footnotesize }\\
{\footnotesize -- Jean-Claude Trichet, President of the ECB, Frankfurt
am Main, 18/11/2010}{\footnotesize \par}

\end{widePar}

\section{Introduction}

Macroprudential oversight refers to surveillance and supervision of
the financial system as a whole. As can be exemplified by recently
founded supervisory bodies with the mandate of safeguarding financial
stability, a system-wide perspective to financial supervision is currently
being accepted and implemented as a common objective of governmental
authorities and supervisors. To this end, the ESRB in Europe, the
FPC in the UK, and the FSOC in the US were founded in the aftermath
of the financial instabilities of 2007\textminus{}2008. Beyond the
soar in availability and precision of data, the transition from firm-centric
to system-wide supervision imposes obvious data needs when analyzing
a large number of entities and their constituents as a whole (see
e.g. \citealp{FloodMendelowitz2009}). As central tasks ought to be
timely and accurate measurement of systemic risks, big data and analytical
models and tools become a necessity. While analytics might aid in
automated modeling, one approach to dealing with complex data and
modeling problems is to improve end users' understanding of them in
order to tap into their expertise. As above noted by Jean-Claude Trichet,
we need means that support \textit{\emph{disciplined and structured
judgmental analysis}} based upon policymakers' experience and domain
intelligence. Further, the mandates of macroprudential supervisors
have to date been stressing or even limited to communication, issuing
warnings and giving recommendations, which boils down to an emphasis
on broad and effective communication of timely information related
to systemic risks.

Financial systems, described by the three pillars of financial intermediaries,
markets and infrastructures, have been shown to be recurringly unstable
due to limitations related to market imperfections \citep{deBandtHartmann2002,Carletti2008}.
Underlying systemic risk, while having no unanimous definition, has
commonly been distinguished into three categories \citep{deBandt2009,ECB2009}:
(\emph{i}) build-up of widespread imbalances, (\emph{ii}) exogenous
aggregate shocks, and (\emph{iii}) spillover and contagion. With the
aim of mitigating system-wide risks, macroprudential oversight is
commonly comprised into a process, where key tasks include (\emph{i})
risk identification, (\emph{ii}) risk assessment, and (\emph{iii})
policy assessment, implementation and follow-up. As a soft policy
intervention, risk communication concerns the overall task of spreading
broadly and effectively timely information related to systemic risks,
as well as other vulnerabilities concerning the financial system and
its macro-financial environment. Fortunately, policymakers and regulators
have access to a broad toolbox of analytical models to measure and
analyze system-wide threats to financial stability. The tasks of these
tools can be mapped to the above listed three forms of systemic risk
(e.g., \citet{ECB2010}): (\emph{i}) early warning of the build-up
of widespread vulnerabilities and imbalances, (\emph{ii}) stress-testing
the resilience of the financial system to a wide variety of exogenous
aggregate shocks, and (\emph{iii}) modeling contagion and spillover
to assess how resilient the financial system is to cross-sectional
transmission of financial instability. While the first aids in risk
identification, the second and third approaches provide means for
risk assessment. Yet, this points out a mismatch between the current
objectives and needs and the available tools: while a key task is
the communication of risks, the toolbox of analytical models lacks
a focus on approaches that support human understanding.

The term visualization has a wide meaning and relates to a number
of interdisciplinary topics, in particular information visualization
and visual analytics. The rationale behind the use of visual representations
and their usefulness relates to traits of the human visual system
(see, e.g., \citet{Ware2004}). \citet{Cardetal1999} assert visualization
as a type of cognitive support or amplification, which leads to a
focus on strengths and weaknesses of human perception. This highlights
the importance of principles for designing visuals that meet the demands
of the human visual system. Although the computer age has brought
visuals, and even the design of them, to the desks of ordinary people,
including policymakers, the most influential literature on data graphics
design still today dates back to work by \citet{Tufte1983} and \citet{Bertin1983}.
Rather than an exact theory, Tufte and Bertin provide a set of principles
and rules of thumb to follow. Techniques supporting visualization
can be divided into two types: graphical representations of data and
means for interaction. While the former can be summarized in various
categories of visualization techniques, such as per output and data,
the latter refer to how the user can interact with or manipulate the
displayed data, such as zooming or panning, which often has its basis
in one or more graphical displays for enabling more fl{}exibility
to explore data. This invokes two questions: \emph{1. (2. how) would
tapping into visualization support risk communication in macroprudential
oversight?}

This paper discusses the role of visualization in macroprudential
oversight at large, especially for the purpose of risk communication.
Risk communication comprises two tasks. Internal communication concerns
spreading information about systemic risks within but at various levels
of the organization, such as among divisions, groups or analysts,
whereas external communication refers to the task of disseminating
information about systemic risks to the general public. In this paper,
we mainly focus on the background and theory of information visualization
and visual analytics, as well as techniques provided within these
disciplines, as potential means for risk communication. The topic
of visualization is in this paper discussed from three viewpoints.
First, based upon the needs for internal and external risk communication,
we define the task of visualization in macroprudential oversight.
Second, we present the so-called macroprudential data cube, by discussing
the type of available data for identifying and assessing systemic
risk, including their structure and its potential implications for
analysis and visualization. Third, we review the current state of
the art in visualization techniques applied to the analysis of systemic
risk. This provides an overview of which tasks should be supported
by visualization and the underlying data to be visualized. Eventually,
the discussion boils down to two essential, but to date rare, features
for supporting the analysis of big financial data and the communication
of risks: analytical visualizations and interactive interfaces.

For visualizing the macroprudential data cube through analytical and
interactive visualization, we provide the VisRisk platform with three
modules: \emph{plots}, \emph{maps} and \emph{networks}.%
\footnote{The VisRisk platform for interactive and analytical applications can
be found here:

\href{http://vis.risklab.fi/}{http://vis.risklab.fi/}%
} \emph{Plots} focus on interactive interfaces for representing large
amounts of data, but do not make use of analytical techniques for
reducing complexity. While \emph{maps} provide analytical means for
representing the three standard dimensions of a data cube in simple
formats, \emph{networks} aim at visualization of the fourth data cube
dimension of interlinkages. As VisRisk enables and is open to the
visualization of any data from a macroprudential data cube, we aim
at providing a basis with which systemic risk indicators and models
can be widely communicated. It is herein illustrated with five web-based
interactive visualizations of systemic risk indicators and models,
of which three make use of analytical visualizations. First, we make
use of analytical techniques for data and dimension reduction to explore
high-dimensional systemic risk indicators and time-varying networks
of linkages. Second, this paper adds interactivity to not only dashboards
of standard risk indicators and early-warning models, but also to
the analytical applications. The ultimate aim of VisRisk, and this
paper at large, is to provide a basis for the use of visualization
techniques, especially those including analytical and interactive
features, in macroprudential oversight in general and risk communication
in particular.

The present paper is organized as follows. While Section 2 discusses
macroprudential oversight and risk communication, Section 3 focuses
on information visualization and visual analytics. In Section 4, we
present an overview of visualization techniques in risk communication
and macroprudential oversight and the macroprudential data cube. Section
5 introduces VisRisk as a general platform for visualizing the macroprudential
data cube, and illustrates it with five web-based interactive visualizations
of systemic risk indicators and models, of which three make use of
analytical visualizations. Section 6 concludes.

\section{Macroprudential oversight and risk communication\label{sec:Macroprudential-oversight}}

Since the date when the still ongoing global financial crisis broke
out, the notion of a macroprudential approach to safeguarding financial
stability has grown consensus among the academic and policymaking
communities alike. Yet, it is by no means a new concept. The \ac{BIS}
applied the term to describe a system-wide orientation of regulatory
frameworks already in the 1970s, and the term appeared in publicly
available material in the mid-1980s (see, e.g., \citet{BIS1986} and
\citet{Borio2011}). The series of recently established macroprudential
supervisory bodies obviously also motivates understanding and disentangling
their specific tasks and functions. 

This section attempts to provide a holistic view of a so-called macroprudential
oversight process. The starting point ought to be acknowledging the
existence of underlying market imperfections, which might cause systemic
risks in the financial system. Then, this section discusses the tasks
within the process, and relates the concept of risk communication
to macroprudential oversight.

\subsection{Market imperfections and systemic risk}

The overall need for macroprudential oversight is motivated by the
potential for systemic risks that threaten the stability of the financial
system. Yet, any form of systemic risk is most often preceded at an
early stage by various market imperfections.%
\footnote{Imperfections in markets may, for instance, take the form of asymmetric
and incomplete information, externalities and public-good characteristics
and incomplete markets, and are to some extent present in most sectors
of the economy.%
} The imperfections, when being related to a financial sector, may
lead to significant fragility of not only individual entities, but
also the entire system \citep{Carletti2008,ECB2009}.%
\footnote{\citet{Carletti2008} relates market imperfections to the financial
sector with a number of examples, such as banks being exposed to deposit
runs due to the maturity transformation by investing short-term deposits
in long-term assets and informational asymmetries between depositors
and borrowers, as well as debtholders and firm managers having so-called
misaligned principal-agency problems, leading to agents not acting
in the best interest of the principal. Likewise, parallels can be
drawn between the public good of financial stability and externalities
like pollution, as each entity manages its own risks with no need
to consider its impact on the system-wide risk as a whole.%
} \citet{deBandtHartmann2002} relate fragilities in financial systems
to three causes: (\emph{i}) strong information intensity and intertemporal
nature of financial contracts and transactions; (\emph{ii}) balance-sheet
structures of financial intermediaries with a high reliance on debts
or leverage, and maturity mismatches between assets and liabilities;
and (\emph{iii}) high degree of interconnectedness between financial
intermediaries and markets. Thus, the underlying market imperfections
may at a later stage propagate as possible systemic risks to financial
stability, which supports the role of governments and other supervisory
authorities in addressing and monitoring systemic risks.

Whereas the above discussion related to the causes of systemic risk,
we have still not defined, and in particular not disentangled, the
types of risk that may occur. While there is no consensus on the definition,
a number of works have categorized systemic risk into three types
(\citealp{deBandt2009,ECB2009,Trichet2009,ECB2010}): \emph{i}) endogenous
build-up and unraveling of widespread imbalances; \emph{ii}) exogenous
aggregate shocks; and \emph{iii}) contagion and spillover.

The first form of systemic risk refers to the risk that \emph{widespread
imbalances}, which have \emph{built up} over time, \emph{unravel abruptly}.
The underlying problems are caused by an endogenous build-up of imbalances
in one or several parts of a fi{}nancial system, such as high concentrations
of lending in certain parts of the economy or credit booms in general,
due to which a shock leading to a repricing of risk may be triggered
by even a small event or change in expectations. This resembles Kindleberger\textquoteright{}s
\citeyearpar{Kindleberger1996} and Minsky\textquoteright{}s \citeyearpar{Minsky1982}
financial fragility view of a boom-bust credit or asset cycle. Second,
systemic risk may also refer to a widespread \emph{exogenous aggregate
shock} that has negative systematic effects on one or many financial
intermediaries and markets at the same time. For instance, if banks
go bad during recessions, they can be said to be vulnerable to economic
downturns. The third form of systemic risk is \emph{contagion and
spillover}, which usually refers to an idiosyncratic problem that
spreads in a sequential fashion in the cross section. For instance,
a failure of one financial intermediary causing the failure of another
financial intermediary given no common risks. These systemic risks
set a need for analytical tools and models for financial stability
surveillance.

\subsection{Analytical tools for financial stability surveillance\label{sub:Analytical-tools-for-macropru}}

The macroprudential approach accentuates the need for a thorough understanding
of not only financial entities, be they instruments, economies, markets
or institutions, but also their interconnections, interlinkages and
system-wide importance. For macroprudential oversight, policymakers
and supervisors need to have access to a broad toolbox of methods
and models to measure and analyze system-wide threats to financial
stability. These analytical tools provide means for two types of tasks:
(\emph{i}) early identification of vulnerabilities and risks, as well
as their triggers, and (\emph{ii}) early assessment of transmission
channels of and a system's resilience to shocks, and potential severity
of the risk materialization.

The literature, while in many aspects being in its infancy, has provided
a variety of tools for financial stability surveillance. Following
\citet{ECB2010}, we can distinguish the models into three broad analytical
approaches that match the identified forms of systemic risks (\emph{the
addressed risks in parenthesis}): (\emph{i}) early-warning models
(\emph{build-up of widespread imbalances}), (\emph{ii}) macro stress-testing
models (\emph{exogenous aggregate shocks}), and (\emph{iii}) contagion
and spillover models (\emph{contagion and spillover}).

First, early-warning models can be used to derive probabilities of
impending systemic financial crises. Second, macro stress-testing
models provide means to assess the resilience of the financial system
to a wide variety of aggregate shocks. Third, contagion and spillover
models can be employed to assess how resilient the financial system
is to cross-sectional transmission of financial instability. 

In addition to models for signaling systemic risks at an early stage,
the literature has provided a large set of coincident indicators that
measure the current state of instability in the fi{}nancial system.
While these serve as means to measure the contemporaneous level of
systemic risk, and thus may be used to identify and signal heightened
stress, they are not designed to have predictive capabilities. This
is not the core focus of this paper, but it is worth noting that \emph{ex
post} measures may serve a function in communicating the occurrence
of unusual events to resolve fear and uncertainty, e.g., after the
so-called flash crash of May 6, 2010 in the \ac{US} \citep{Bisiasetal2012}.
In the sequel of this section, we focus on the three analytical approaches
to derive tools for early identification and assessment of risks,
as well as their precise use in macroprudential oversight.

\subsection{The macroprudential oversight process\label{sub:A-framework-for-macroprudential-oversight}}

\begin{figure}
\begin{centering}
\includegraphics[width=1\columnwidth]{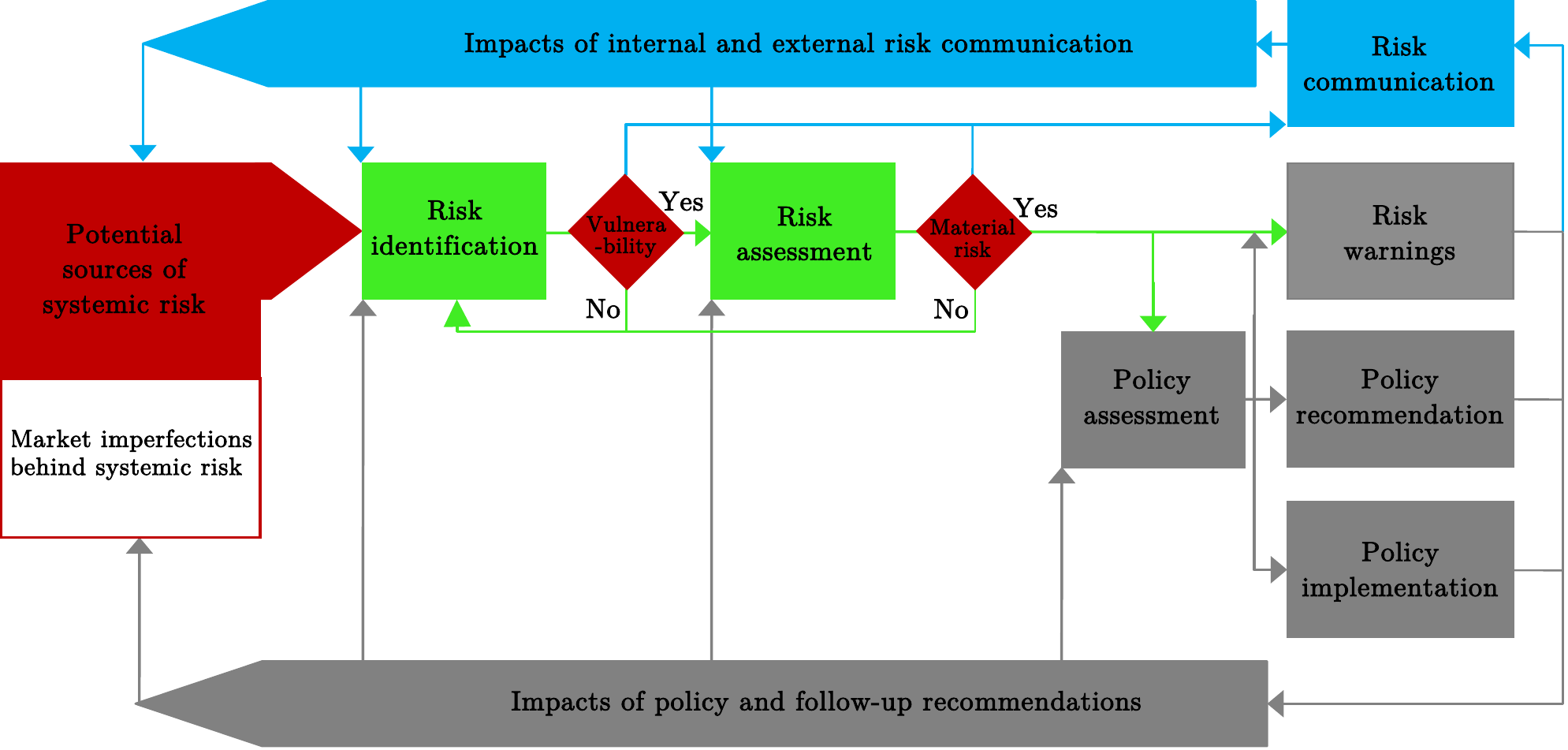}
\par\end{centering}

\textbf{\scriptsize Notes}{\scriptsize : The figure illustrates the
macroprudential oversight process. The }\textcolor{red}{\scriptsize red
components}{\scriptsize{} represent risks and vulnerabilities and the
}\textcolor{green}{\scriptsize green components}{\scriptsize{} represent
the need for risk identification and assessment, }\textcolor{yellow}{\scriptsize gray
components}{\scriptsize{} represent policy assessment and implementation,
as well as risk warnings, policy recommendations and to follow-up,
and }\textcolor{blue}{\scriptsize blue components}{\scriptsize{} represent
overall risk communication. The figure is adapted from \citet{ECB2010}.}{\scriptsize \par}

\centering{}\caption{\label{Fig:oversight} The macroprudential oversight process.}
\end{figure}

The above described market imperfections, and thereby caused systemic
risks, are a premise for macroprudential oversight. Accordingly, the
above three analytical approaches aim at signaling these systemic
risks at an early stage. In terms of a process, Figure \ref{Fig:oversight}
puts forward the steps of the process that a macroprudential supervisory
body follows.%
\footnote{A macroprudential supervisory body is an institution tasked with macroprudential
oversight of the financial system and the mandate of safeguarding
financial stability. Examples are the European Systemic Risk Board
in Europe, the Financial Policy Committee in the \ac{UK}, and the
Financial Stability Oversight Council in the \ac{US}.%
} As described in \citet{ECB2010}, macroprudential oversight can be
related to three steps: \emph{i}) risk identification, \emph{ii})
risk assessment, and \emph{iii}) policy assessment, implementation
and follow-up, as well as giving risk warnings and policy recommendations.
The process in Figure \ref{Fig:oversight} deviates from that in \citet{ECB2010}
by explicitly introducing a fourth task of risk communication and
its feedback loop. In the figure, \textcolor{red}{red components}
represent risks and vulnerabilities, \textcolor{green}{green components}
represent the need for risk identification and assessment, \textcolor{yellow}{gray
components} represent policy assessment and implementation, as well
as risk warnings, policy recommendations and follow-up, and \textcolor{blue}{blue
components} represent overall risk communication. Moreover, following
\citet{Buba2013}, we can distinguish the final instruments into different
levels of organization: \emph{i}) soft (communication), intermediate
(warnings and recommendations) and hard (interventions).

In the \emph{first step} of the supervisory process, the key focus
is on identifying risks to stability and potential sources of vulnerability.
The vulnerabilities and risks could exist in any of the three pillars
of the financial system: financial intermediaries, financial markets
and financial infrastructure. The necessary analytical tools to identify
possible risks, vulnerabilities and triggers come from the set of
early-warning models and indicators, combined with the use of market
intelligence, and expert judgment and experience. This involves ranking
risks and vulnerabilities as per intensity, as well as for assigning
probabilities to specific shocks or future systemic events. 

In the \emph{second step} of the process, the rankings and probabilities
may be used to assess the identified risks. Beyond market intelligence,
as well as expert judgment and experience, risk assessment makes use
of analytical tools mainly from the set of macro stress-testing models
and contagion models. In macro stress-testing, simulations of most
plausible risk scenarios show the degree of impact severity on the
overall financial system, as well as its components. Contagion models,
on the other hand, might be used through counterfactual simulations
to assess the impact of specific failures on the entire financial
system and individual institutions. The first and the second step
of the process should not only provide a list of risks ordered according
to possible severity, but also contain their materialization probabilities,
losses given their materialization, and real losses in output and
welfare, as well as their possible systemic impact. Hence, these two
initial steps in the process aim at early risk identification and
assessment and provide means for safeguarding financial stability.

The \emph{third step} of the process involves the assessment, recommendation
and implementation of policy actions as early preventive measures.
Based upon the identified and assessed risks, a macroprudential supervisory
body can consider giving a wide variety of risk warnings and recommendations
for other parties to use policy instruments, as well as implementations
of policies given the instruments at hand. To steer their decisions,
the policy assessment step can make use of the same analytical tools
used for risk identification and assessment. Likewise, analytical
tools may support assessment prior to issuing risk warnings and giving
policy recommendations. While the use of policy tools is beyond the
mandate of some macroprudential supervisory bodies, actions tailored
to the needs of a system-wide orientation are becoming a key part
of financial regulation and supervision.%
\footnote{For instance, as interest rate policy may be a too blunt and powerful
tool with material impact on other parts of the economy, the policies
could take the form of tighter standards -- e.g., requirements on
capital adequacy, provisioning, leverage ratios, and liquidity management
-- for individual financial institutions with larger contributions
to systemic risk and calibrated to address common exposures and joint
failures. Macroprudential regulation and tools may also be used for
accumulating buffers or reserves in good economic times to be used
during worse times.%
} As illustrated in Figure \ref{Fig:oversight}, policies have an impact
on not only the assessment of policy and identification and assessment
of risks, but obviously also directly on market imperfections and
the accumulation of systemic risks.

The \emph{fourth step}, while not always being the last task to be
performed, concerns risk communication, and its own feedback loop,
which is a central part of this paper and is thus the topic of the
following subsection.

\subsection{Risk communication }

The above discussion untangled overall risk communication as a separate
step in the macroprudential oversight process. Although the tasks
of overall risk communication is inherently different, a integral
part of warnings, and policy recommendations and implementations also
makes use of the communication channel, in which the overall task
concerns disseminating information. The above subsection positioned
risk communication within the macroprudential oversight process, yet
did not provide a detailed discussion. This subsection brings up the
role and possible forms of risk communication.

\begin{figure}
\begin{centering}
\includegraphics[width=1\columnwidth]{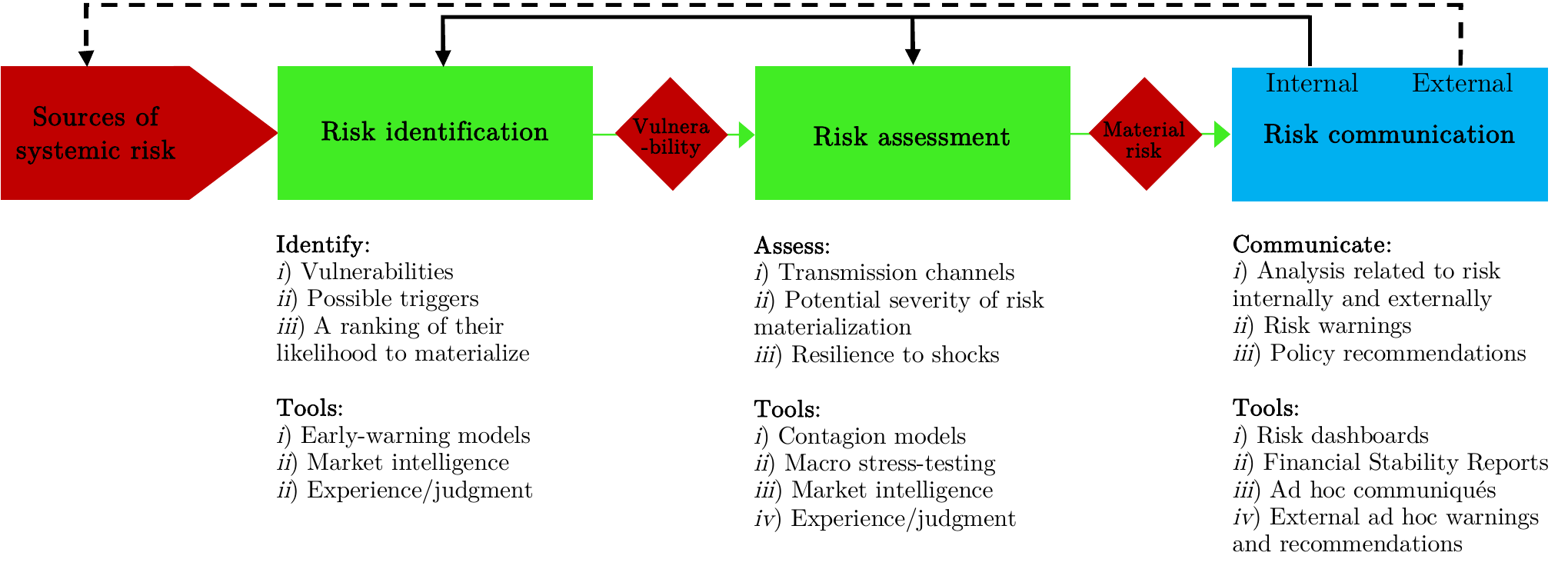}
\par\end{centering}

\centering{}\caption{\label{Fig:risk_communication} Risk communication in macroprudential
oversight.}
\end{figure}

Risk communication describes the task of disseminating broadly and
effectively timely information related to systemic risks and other
vulnerabilities of the pillars of the financial system. Moreover,
macro-financial imbalances and risks at the sector level (i.e., household,
corporate, foreign and government). From the viewpoint of risk communication
of a macroprudential supervisory body, Figure \ref{Fig:risk_communication}
simplifies the macroprudential oversight process into three key steps:
risk identification, risk assessment and risk communication. Hence,
it concentrates on the soft type of intervention. Following the discussion
thus far, the figure also summarizes the key tasks and available tools
in each of the process steps. Building upon Figure \ref{Fig:oversight},
where risk communication was shown to feed into both underlying systemic
risks and the tasks of risk identification and assessment, Figure
\ref{Fig:risk_communication} disentangles the two types of feedback
depending on whether one communicates internally or externally. Internal
communication concerns spreading information about systemic risks
within, but at various levels of, the organization, such as among
divisions, groups or analysts, whereas external communication refers
to the task of disseminating information about systemic risks to the
general public. As shown in Figure \ref{Fig:oversight}, it is worth
to note that the information to be communicated might derive directly
from the risk identification or assessment steps or then feed back
only after recommendations, warnings and implementations in step three.

\emph{Internal communication} refers to a range of activities at different
levels of the organization. Despite that communication of macroprudential
supervisors like central banks commonly refers to disseminating information
externally, \citet{Mohan2009} stresses that they, particularly policy
and research wings, ought to pay attention to and play a pro-active
role in the dissemination and effective communication of policy and
analysis to internal staff. Borrowing from \citet{Mitchell2001},
this necessitates to \emph{\textquotedblleft{}sell the brand inside\textquotedblright{}}.
This is further exemplified by Mohan, as \emph{''internal communication
could act as a conduit for external communication''}, where staff
oftentimes play a key role as ambassadors of the institution.

We relate this to three levels of organization. At the lowest level,
it relates to the use of means for communicating information to support
individual analysts, which in the information age mainly relates to
human-computer interaction (see, e.g., \citet{Dixetal2004}). In this
case, communication relates to the interaction between a computer
and interface with an analyst or policymaker overall. The second level
of organization concerns the task of communicating among and within
smaller groups, teams or divisions. Within the limits of a small,
specialized audience, this involves not only supporting knowledge
crystallization, but also supporting the dissemination of knowledge.
While the former relates to communication within groups of analysts
and policymakers, the latter concerns communicating the results or
insights of one member or team to another. At the final, third level
of an organization, communication relates to disseminating the information
gathered within the specialized teams, groups or divisions to high-level
management or the rest of the organization. As the audience becomes
broader, the means for disseminating information, as well as the information
itself, need also to be of a more general nature for the message to
be digestible. At each of these levels, while the means for internal
communication are various communiqués and announcements, a policymaker
can tap into the tools and expertise, such as analytical models, market
intelligence and expert judgment and experience, available at the
supervisory body.

Beyond stressing the importance of it, \citet{Mohan2009} pinpoints
four suggestions that support internal communication: (\emph{i}) arranging
internal seminars in all regional offices and training colleges after
external policy announcements and report publications (\emph{ii})
the publication of FAQs (frequently asked questions) on the internet/intranet
on each policy matter, as well as the provision of educational resources,
(\emph{iii}) the publication of a working paper series with a clear
disclaimer that ascribes research findings and overall opinions to
the authors, and (\emph{iv}) internal notes of various divisions should
be made available internally, as they are of immense analytical value.

\emph{External communication }refers to conveying information about
systemic risks to the general public, including other authorities
with responsibility for financial stability and overall financial-market
participants, such as laymen, professional investors and financial
intermediaries. An obvious difference in relation to internal communication
relates to a comparatively larger heterogeneity in the audience. Yet,
while a voluminous literature supports internal communication within
organizations, one needs to look to a different direction in order
to answer why a macroprudential supervisory body ought to communicate
risk externally. Paraphrasing the role of communication in monetary
policy, \citet{Bornetal2013} argue that communication aims at (\emph{i})
improving credibility of central banks (relating communication to
transparency and reputational purposes), (\emph{ii}) enhancing effectiveness
of policy (relating communication to financial stability contributions),
and (\emph{iii}) to make central banks accountable (relating to explicit
communication of identified risks and vulnerabilities). 

By means of an example, a key means for such communication is through
quarterly or biannual Financial Stability Reports, a recent phenomenon
that has quickly spread to a large number of central banks. With the
aim of understanding the overall purpose of communication, a survey
among central bankers by \citet{Oosterlooa2004} pinpoints three main
reasons for publishing Financial Stability Reports: (\emph{i}) to
contribute to overall financial stability, (\emph{ii}) to increase
the transparency and accountability, and (\emph{iii}) to strengthen
co-operation between authorities with financial stability tasks. At
the same time, \citet{Allenetal2004} were consulted to study the
effectiveness of communication related to financial stability at Sveriges
Riksbank, the first central bank to publish a Financial Stability
Report in 1997. Their external evaluation focused on Financial Stability
Reports in 2003 and overall analytical framework and other work on
financial stability. This resulted in ten recommendations, such as
having and making objectives explicit and precise in the Financial
Stability Reports, also covering other sectors than banks, such as
the insurance sector, and making charts and underlying data easily
downloadable. Although the case study focuses solely on the Riksbank,
its conclusion of \emph{''the Riksbank is doing a very good job in
fulfilling its financial stability responsibilities''} feels justified,
as they were indeed forerunners in the tasks at that point in time.
Further, \citet{Cihak2006} systematically reviews Financial Stability
Reports, and documents a considerable increase in sophistication over
time and improvements in not only their scope, but also the analytical
tools with which analysis is conducted.

In a recent study focusing on communication, the overall finding by
\citet{Bornetal2013} was that Financial Stability Reports, as well
as ad hoc speeches and interviews, affect financial markets by creating
news (i.e., co-occurring jumps in stock returns) and reducing noise
(i.e., decreasing market volatility). Further, \citet{Ekholm2012}
-- the Deputy Governor of the Riksbank -- notes that there is a strive
for not only openness and transparency, but also clear external communication.
In particular, Ekholm notes that during times of crisis ''\emph{a
\textquotedblleft{}negative\textquotedblright{} but reliable announcement
can {[}...{]} be better for confidence than a \textquotedblleft{}positive\textquotedblright{}
but uncertain announcement}''. Along these lines, the means for external
communication concern the use of not only Financial Stability Reports
published at regular intervals, but also risk warnings and recommendations
communicated through various ad hoc public announcements. A recent
addition to the toolbox of communication approaches is the publication
of a risk dashboard, which essentially involves developing and publishing
a set of indicators to identify and measure systemic risk.%
\footnote{While the risk dashboard of the European Systemic Risk Board has been
published since September 2012, the European Banking Authority published
its first risk dashboard in October 2013.%
} Like in internal communication, a policymaker communicating externally
also ought to tap into not only analytical models and tools at hand,
but also market intelligence and expert judgment and experience when
representing and judging the current state of risks and vulnerabilities.
The latter becomes an obvious input when drafting any types of textual
policy communiqués.

Thus far, we have taken the analytical models as given -- both those
used in risk identification and assessment and those used in risk
communication. Yet, whereas analytical tools have clearly been designed
to address the tasks in risk identification and assessment, they are
in no, or little, explicit focus in the task of risk communication.
In particular, there is a clear lack of integration of tools for the
common objective of a macroprudential supervisory body, whose one
key focus is to communicate identified and assessed risks. This paper
asks the question: \emph{is there something to be gained by tapping
into the fields of information visualization and visual analytics
when communicating systemic risk?}

\section{Information visualization and visual analytics}

The visualization of complex data has lately emerged as one of the
key aids to support \ac{EDA}, though the task of EDA dates back to
Tukey's early work in the 1970s (e.g., \citet{Tukey1977}). Whereas
advanced visual representations of data are common in a wide range
of disciplines and domains, the use of these types of representations
are rare in the communication of macroprudential bodies or supervisory
authorities at large. The key aim of this section is to discuss the
rationale behind the usefulness of visual representations, how visuals
should be designed to meet the demands of the human visual system
and categorizations of approaches to visualization. At a higher level,
this section covers the discipline of information visualization --
and the more recent derivative, visual analytics -- in order to support
a later discussion of their merits in macroprudential oversight.

Information visualization as a discipline has its origin in the fields
of human-computer interaction, computer science, graphics and visual
design. A more precise definition of it is \emph{\textquotedblleft{}the
use of computer-supported, interactive, visual representations of
abstract data to amplify cognition\textquotedblright{}} \citep{Cardetal1999},
which highlights improving human understanding of data with graphical
presentations or graphics. Tools for information visualization are
mainly and best applied for \ac{EDA} tasks, and most commonly aim
at browsing a large space of information. While being in a highly
common and general setting, \citet{Lin1997} lists browsing to be
useful when: 
\begin{enumerate}
\item there is a good underlying structure and when related items can be
located close by;
\item users are unfamiliar with the contents of the collection;
\item users have little understanding of the organization of a system and
prefer to use a method of exploration with a low cognitive load;
\item users have difficulty in articulating or verbalizing the specific
information need; and
\item users search for information that is easier to recognize than describe.
\end{enumerate}
The above list relates to situations when visualization is useful,
yet we still need to discuss the elements of information visualization
in-depth. The rest of this section focuses on three subtopics of information
visualization -- human perception and cognition, data graphics design
and visualization techniques -- in order to end with a discussion
of visual analytics, which combines analytical models with visual
representations.

\subsection{Human perception and cognition}

Much attention is given to the design of visual representations. While
being important, a discussion about information visualization cannot
start from the colors, shapes and other features used for representing
data. Instead, a starting point to visual communication ought to be
to understand and acknowledge the capabilities and limits of the human
information and visual system. The visual system comprises the human
eye and brain, and can be seen as an efficient parallel processor
with advanced pattern recognition capabilities (see, e.g., \citet{Ware2004}).
The focus of human perception is the understanding of sensory information,
where the most important form is visual perception. The final \acl{IA}
of information visualization can be viewed as a type of cognitive
support. The mechanisms of cognitive support are, however, multiple.
Hence, visualization tools should be targeted to exploit advantages
of human perception.

Mostly, arguments about the properties and perception capabilities
of the human visual system rely on two grounds: \emph{i}) information
theory \citep{ShannonWeaver1963}, and \emph{ii}) psychological fi{}ndings.
\emph{Information theory} states that the visual canal is best suited
to carry information to the brain as it is the sense that has the
largest bandwidth. \citet{Ware2004} asserts that there are two main
\emph{psychological theories} for explaining how to use vision to
perceive various features and shapes: preattentive processing theory
\citep{Triesman1985} and gestalt theory \citep{Koffa1935}. Prior
to focused attention, preattentive processing theory relates to simple
visual features that can be perceived rapidly and accurately and processed
effectively at the low level of the visual system. Whereas more complex
visual features require a much longer process of sequential scanning,
preattentive processing is useful in information visualization as
it enables rapid dissemination of the most relevant visual queries
through the use of suitable visual features, such as line orientation,
line length or width, closure, curvature and color \citep{Feketeetal2008}.
At a higher cognitive level, gestalt theory asserts that our brain
and visual system follow a number of principles when attempting to
interpret and comprehend visuals. \citet{Ware2004} summarizes the
principles as follows:

\vspace{0.2cm}

\textbf{Proximity:} Items close together are perceptually grouped
together.

\textbf{Similarity:} Elements of similar form tend to be grouped together.

\textbf{Continuity:} Connected or continuous visual elements tend
to be grouped.

\textbf{Symmetry:} Symmetrical elements are perceived as belonging
together.

\textbf{Closure:} Closed contours tend to be seen as objects.

\textbf{Relative size:} Smaller components of a pattern tend to be
perceived as objects.

\vspace{0.2cm}

The principles of gestalt theory can easily be related to some more
practical concepts. For instance, most projection methods, when aiming
at visualizing data, may be seen to relate to the proximity principle,
as they locate high-dimensional data with high proximity close to
each other on a low-dimensional display, whereas others are pushed
far away. Likewise, a time trajectory may be paired with continuity.
More related to the cognition of visualization, \citet{Feketeetal2008}
relate the core benefi{}t of visuals to their functioning as a frame
of reference or temporary storage for human cognitive processes. They
assert that visuals are external cognition aids in that they augment
human memory, and thus enable allocating a larger working set for
thinking and analysis. In the above stated definition of information
visualization by \citet{Cardetal1999}, visuals are presented as a
means to \textquotedblleft{}amplify cognition''. The same authors
also list a number of ways how well-perceived visuals could amplify
cognition:
\begin{enumerate}
\item by increasing available memory and processing resources;
\item by reducing the search for information;
\item by enhancing the detection of patterns and enabling perceptual inference
operations;
\item by enabling and aiding the use of perceptual attention mechanisms
for monitoring; and
\item by encoding the information in an interactive medium.
\end{enumerate}
Not to disturb legibility of this section, examples of the five ways
to amplify cognition are given Appendix A.1. Yet, while visualization
provides ample means to amplify cognition, it is also worth looking
into matters concerning human perception and cognition that may hinder,
disturb or otherwise negatively affect how visualizations are read.
An essential part of visualization is to take into account the deficiencies
and limitations of human perception. Again, to keep this section shorter,
more detailed exemplifications are provided in Appendix A.2. Accordingly,
an understanding of the functioning of the human visual system aids
in producing effective displays of information, where emphasis is
on presented data such that the patterns are likely to be correctly
perceived.

\subsection{Data graphics design}

Based upon features of the human visual system, and avenues for supporting
perception and cognition, the literature on data graphics design has
its focus on the principles for visual representations of data. Herein,
the focus is on the early, yet still today influential, work by \citet{Tufte1983}
and \citet{Bertin1983}. Their works, while being principles for graphics
design, are also valid for overall computer-based visualization. Tufte's
set of principles are called a theory of data graphics, whereas Bertin\textquoteright{}s
work is most often denoted a framework of the planar and retinal variables.
Yet, rather than an exact theory, Tufte and Bertin provide a set of
rules of thumb to follow.

The following overview is included to provide concrete guidelines,
in addition to the higher-level discussion of perception and cognition.
Herein, we will only focus on the key components of frameworks and
theories by Bertin and Tufte. We start from Bertin's \citeyearpar{Bertin1983}
framework called the Properties of the Graphic System, which consists
of two planar and six retinal variables. The two planar variables
are the $x$ and $y$ dimensions of a visual, whereas the six retinal
variables describe the following visual marks on the plane: size,
value, texture, color, orientation and shape. The eight variables
can be categorized according to the following levels of organization,
or so-called perceptual properties:
\begin{enumerate}
\item \textbf{Associative} $(\equiv)$: If elements can be isolated as belonging
to the same category, but still do not affect visibility of other
variables and can be ignored with no effort.
\item \textbf{Selective} $(\neq)$: If elements can immediately and effortlessly
be grouped into a category, and formed into families, differentiated
by this variable, whereas the grouping cannot be ignored.
\item \textbf{Ordered} $(\mathbf{O})$: If elements can perceptually be
ordinally ranked based upon one visually varying characteristic.
\item \textbf{Quantitative} $(\mathbf{Q})$: If the degree of variation
between elements can perceptually be quantified based upon one visually
varying characteristic.
\end{enumerate}
When having an understanding of the four levels of organization, we
can return to Bertin's \citeyearpar{Bertin1983} eight visual variables.
We refer to Appendix A.3 for an in-depth discussion of the variables,
as we herein only provide a summary of the variable properties. Bertin
describes the plane, and its two dimensions $(x,y)$, as the richest
variables. They fulfill the criteria for all levels of organization
by being associative $(\equiv)$, selective $(\neq)$, ordered $(\mathbf{O})$
and quantitative $(\mathbf{Q})$. The retinal variables, on the other
hand, are always positioned on the plane, and can make use of three
types of implantation: a point, line, or area. Their perceptual properties
are as follows: size $(\neq,\mathbf{O},\mathbf{Q})$, value $(\neq,\mathbf{O})$,
texture $(\equiv,\neq,\mathbf{O})$, color $(\equiv,\neq)$, orientation
$(\equiv,\neq)$, and shape $(\equiv)$.

A complement to Bertin's framework is the Theory of Data Graphics
by Tufte \citeyearpar{Tufte1983}, which consists of a large number
of guidelines for designing data graphics. The two key, broad principles
are graphical excellence and graphical integrity.

\citet{Tufte1983} defines \emph{graphical excellence} as a graphic
that ''\emph{gives to the viewer the greatest number of ideas in
the shortest time with the least ink in the smallest space}''. The
principle of graphical excellence summarizes a number of his guidelines
that encourage graphical clarity, precision, and efficiency: \emph{i})
avoid distortions of what the data have to say; \emph{ii}) aid in
thinking about the information rather than the design; \emph{iii})
encourage the eye to compare the data; \emph{iv}) make large data
sets coherent; \emph{v}) present a large number of data in a small
space; \emph{vi}) reveal data at multiple levels of detail ranging
from a broad overview to fine detail; \emph{vii}) and closely integrate
statistical and verbal descriptions of the data. The second of Tufte's
\citeyearpar{Tufte1983} principles, \emph{graphical integrity}, relates
to telling the truth about data. To follow this principle, Tufte provides
six key guidelines: \emph{i}) visual representations of numbers should
be directly proportional to the quantities which the visuals represent;
\emph{ii}) clear and detailed labeling should be used to avoid ambiguity;
\emph{iii}) show data variation, not design variation; \emph{iv})
deflate and standardize units when dealing with monetary values; \emph{v})
the number of dimensions depicted should not exceed the number of
dimensions in data; and \emph{vi}) data should not be showed out of
context. The overall aim of principles related to graphical integrity
is to avoid deception and misinterpretation. This provides a brief
overview of Tufte's rules of thumb, whereas interested readers are
referred to Appendix A.3 as well as the original sources.

Bertin\textquoteright{}s and Tufte's principles provide a guiding
set of rules of thumb to follow when spanning the space of two-dimensional
visualizations. Yet, visualizations, not the least interactive visualizations,
go beyond a static two-dimensional space by including additional visual
variables, such as depth and time. This highlights requirements on
the visualization techniques and tools, where interaction is essential.

\subsection{Visualization techniques and interfaces}

The literature has provided a long list of techniques for creating
visual representations and interfaces, with the aim of supporting
human perception and cognition. This subsection focuses mainly on
a rough overview, as well as a brief and simple taxonomy, of methods,
rather than a detailed survey. Obviously, a key issue of information
visualization is what formats and features the methods will help to
organize and visualize, as well as how that relates to the use of
the capabilities of the human visual system. Techniques supporting
information visualization can be divided into two types: graphical
representations of data and interaction techniques. The former type
refers to the visual form in which the data or model is displayed,
such as standard bar and line charts. Yet, visualization may often
refer to the use of manipulable graphical displays of data. The latter
type of techniques refer to how the user can interact with or manipulate
the graphical displays, such as zooming or panning. These oftentimes
have their basis in one or more graphical displays such that they
enable more freedom and fl{}exibility to explore the data.

From the viewpoint of the underlying data, rather than the formats
of visual displays, \citet{Zhangetal2012} categorize visualization
techniques into four groups: numerical data, textual data, geo-related
data and network data. Yet, a categorization of visualization techniques
as per the types of data does not differentiate all possibilities
of techniques. While being some years old, \citet{KeimKriegel1996}
provide a five-category grouping of techniques by the visualization
output that still today holds merit: geometric, icon-based, pixel-oriented,
hierarchical, and graph-based techniques. In addition, Keim and Kriegel
also illustrate the existence of a wide range of hybrids that make
use of multiple categories. While a description of each category and
examples of techniques can be found in Appendix A.4, it only highlights
the large number and wide variety of available techniques. The categorization
of visualizations as per data and display, while highlighting challenges
in choosing the correct technique for the data and the task at hand,
provides guidance in the choice. For instance, one obvious factor
to define the nature of the chosen visualization technique is the
properties of the data, such as the form of data, dimensionality of
data, data structures and size of data. Further, another factor is
the expected output and purpose of use, such as predictive \emph{vs}.
exploratory, temporal \emph{vs}. cross-sectional, and univariate \emph{vs}.
multivariate analysis and similarity \emph{vs}. dissimilarity matching,
as well as other purposes related to a focus on geo-spatial visualization
and network relationships, for instance. While there obviously is
no one way to choose the correct technique, considering the two dimensions
of data and display, as well as other restrictions, demands and needs
for the task, provides an adequate basis.

Given a technique, a critical factor of information visualization
is, however, the possibility to interact with the visuals. A common
guideline for interactions with visualizations is the visual information
seeking mantra \citep{Shneiderman1996}: ''\emph{Overview first,
zoom and filter, then details-on-demand}''. Whereas \citet{Shneiderman1996}
characterizes the mantra with seven abstract tasks, we focus only
on the following four explicitly mentioned ones: First, a user should
gain an \emph{overview} of the entire collection through a high-level
representation. Second, users should have the possibility to \emph{zoom}
in on a portion of items that are of particular interest. Third, there
should exist the possibility to \emph{filter} out or to eliminate
uninteresting and unwanted items, such as allowing users to specify
which items to display. Fourth, the user should have the option to
select an item or group of items to get further \emph{details-on-demand},
such as clicking a group or individual items to browse descriptive
information. 

This provides a starting point to data visualization and user interaction,
but does still not address the role of analytical techniques in visualization.
The next step is to combine graphical representations of data and
interaction techniques with analytical methods.

\subsection{Visual analytics\label{sub:Visual-analytics}}

A recent, rapidly growing discipline is that of visual analytics.
By adding analytics to the ingredients of information visualization,
we end up with the original definition of visual analytics \citep{ThomasCook2005}:
''\emph{the science of analytical reasoning facilitated by interactive
visual interfaces}''. Hence, the field of visual analytics has strong
roots in information visualization. Likewise, visual analytics is
obviously strongly related to overall data analytics. The term visual
data mining descends from the integration of the user in the data
mining (or analytics) process through visualization techniques and
interaction capabilities (see, e.g., \citet{Keim2001}). This has
taken visual analytics to be applied in areas with challenging problems
that were unsolvable using standalone automatic or visual analysis
(see, e.g., \citet{Keimetal2009}). In particular, while automated
computational processing enables scaling to larger and more challenging
tasks, humans exhibit traits that enable a deeper understanding \citep{Rischetal2008}.
This highlights the importance of coupling the strengths of computational
and human information processing. When also including the interaction
with analytical parameters, visual analytics is not only helpful in
applications involving large, complex data, but also those involving
complex analytical processes requiring monitoring and interaction.

\begin{figure}
\begin{centering}
\includegraphics[width=0.8\columnwidth]{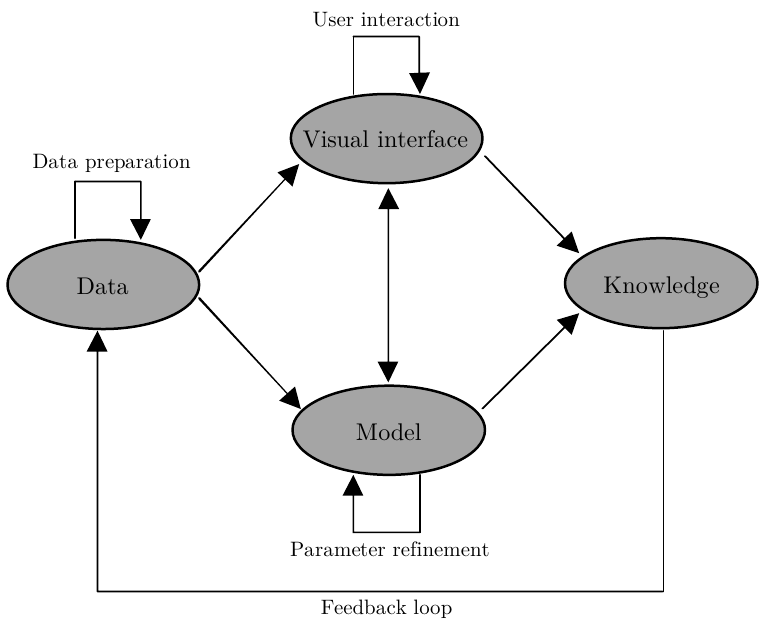}
\par\end{centering}

\noindent \textbf{\scriptsize Notes}{\scriptsize : The figure represents
the visual analytics process. The figure is adapted from \citet{Keimetal2010}.}{\scriptsize \par}

\centering{}\caption{\label{Fig:visualanalytics} Visual analytics process.}
\end{figure}

Since we derive visual analytics from three above presented concepts
-- graphical representations of data, interaction techniques and analytical
techniques -- there is no need to repeat the discussion of each component.
Further, the above presented information seeking mantra only mentions
visualization, yet does not integrate it with analytics. \citet{Keimetal2006}
propose combining an analytics process and the information seeking
mantra for a visual analytics mantra: ''\emph{Analyze first, show
the important, zoom, filter and analyze further, details on demand}''.
The authors exemplify the visual analytics mantra with analysis of
large network security data. As graphical representations of raw data
is infeasible and seldom reveals deep insights, the data need to fi{}rst
be analyzed, such as computing changes and intrusion detection analysis.
Then, the outcome of the automated analysis is visualized. Out of
the displayed results, the user fi{}lters out and zooms in to choose
a suspicious subset of all recorded intrusion incidents for further,
more careful analysis. Thus, the mantra involves automated analysis
before and after the use of interactive visual representations. Following
the mantra, the visual analytics process discussed in \citet{Keimetal2010}
is presented in Figure \ref{Fig:visualanalytics}. The key steps in
the process are data preparation, visual and automatic analysis, and
knowledge consolidation. After the step of data preprocessing and
transformations, the user selects between visual or automatic analysis
methods. The user might prefer to start from whichever of the two
tasks, and might then iterate multiple times in between the two components
of data visualization and interaction and automatic analysis. Finally,
after alternating between visual and automatic methods, the thus far
gained knowledge is not only gathered, but also transferred through
a feedback loop to support future analysis.

To be more general, the viewpoint we take on visual analytics relates
more to human-computer cooperation. Relying on data as an input, this
concerns the combination of various forms of computational and human
processing, in order to support the knowledge crystallization process.
Beyond aids for visualization and interaction, which support the development
of a mental model, this includes advanced analytical techniques as
a means to rigorous formal models. It is the interplay between formal
and mental models that is the basis for knowledge creation, including
tasks ranging from sensemaking to reasoning to decisionmaking.

\section{Visualization in macroprudential oversight}

The discussion thus far has concerned macroprudential oversight, in
particular the role of risk communication, and the visualization of
data, in particular the fields of information visualization and visual
analytics. It is obvious that what follows now is their coupling:
\emph{how can visualization be used to support macroprudential oversight
in general and risk communication in particular?}

This section starts by defining the task of visualization in internal
and external risk communication. Then, we turn to a discussion about
the type of available data for measuring and assessing systemic risk
in macroprudential oversight. Finally, we relate the above discussed
topics in visualization to the tasks of macroprudential oversight,
including current use of various categories of visualizations as per
the three types of systemic risk models.

\subsection{Visual risk communication}

Data visualization can serve multiple purposes in macroprudential
oversight overall and risk communication in particular. Despite a
range of different purposes of use, visual representations can generally
be classified to serve the purpose of communicating information to
two audiences: \emph{i}) internal and \emph{ii}) external.

The purpose of use in \emph{internal communication} relates to enhancing
the understanding of policymakers on various levels. One obvious task
is to support analysts themselves, and within other groups of active
participants in the process of collecting data and deriving indicators
and analytical models. This particularly concerns human-computer interaction,
as visual interfaces are used as interactive tools to better understand
complex phenomena with no one known message to be communicated. Whereas
the more common setting is likely to involve collecting data and deriving
indicators, which could be supported by information visualization,
there is also a focus on analytical approaches and understanding derived
models, which points more towards visual analytics. This provides
input to two purposes: data analysis and decisionmaking. An essential
part of data analysis, particularly predictive analytics, involves
data understanding and preparation, which indeed benefits from the
use of visual interactive interfaces. Likewise, visualizing the output
of data analysis provides ample means to support in making better
decisions. This provides a range of internal communication tasks.

Beyond supporting individuals, one may want to communicate to other
involved parties, for which visuals would be used to communicate a
particular message or result to entire divisions, the the management
and even at the level of the entire organization. At the lower level,
the key task is to provide means for interaction with visuals in order
to amplify cognition, which supports a better understanding and modeling
of the task at hand. As above noted, the case of data analysis by
low-level analysts is a standard setting, and mainly involves the
task of human-computer interaction. In the context of low-level internal
communication of systemic risk modeling, \citet{FloodMendelowitz2009}
note that data exploration is an area where visualization tools can
make a major contribution. They point to the fact that certain tasks
of classification, analysis and triage can be automated, whereas many
require a human analyst, such as the difficulty to train a well-performing
machine to analyze anomalous financial market activity. This follows
the very definition of visual analytics. At the higher level, the
focus is more on reporting and presentation of information by means
of visuals. An example could be the dissemination of identified risks
by a risk identification division for further analysis at a risk assessment
division, or even to the board or president of an organization. Moreover,
disseminating results of analytical models within and among divisions
provides scrutiny, which is likely to either improve model credibility
or quality. Thus, to sum up, a major concern is how results of the
risk identification and assessment tasks are communicated to a wide
range of stakeholders in easily understandable formats, with the ultimate
aim of achieving transparency and accountability at an internal level.

A possible criticism is that visual inspection of complex data leaves
room for human judgment, particularly when used for the support of
decisionmaking. Contrary to the concept of economic ''rationality'',
human adjustment is asserted to make visual analysis prone to a so-called
personal forecast bias, which has been associated with traits like
prejudice, undue conservatism and unfounded optimism, as among others
postulated by \citet{Armstrong1985}: ''Don't trust your common sense.''
Yet, it is more than common sense that every policy decision relies
at least partly on judgment, which might or might not be biased. And
it is also worth noting that the decisions are not made by statistical
models, but rather by humans, who are also eventually accountable
for them. A number of works have, however, shown that judgmental adjustments
to macroeconomic model-based forecasts improve accuracy more often
than not. For instance, \citet{McNees1990} shows that judgmental
adjustments to macroeconomic forecasts resulted in a 15\% improvement
in accuracy.

On a more general note, which points towards all levels of organization,
\citet{Mohan2009} suggests that we can do much more to improve internal
communication due to \emph{''the increasing availability of electronic
communication at low cost''}. Mohan further stresses the importance
of innovative ways as means to accomplish this, as the time of management
is unavoidably limited. Turning to relations to Information Visualization
and Visual Analytics, both can be seen as supportive means for internal
communication. While the former supports spreading knowledge, the
latter has a focus better aligned with creating knowledge.

\emph{External communication}, on the other hand, refers to conveying
information to other authorities with responsibility for financial
stability and overall financial-market participants, such as laymen,
professional investors and financial intermediaries. So, how do visual
means aid in the task? Along the lines of the conclusions in \citet{Bornetal2013},
even though they rely on effects of mainly textual communication,
providing improved means for communication is expected to increase
effectivity and certainty of financial markets, particularly through
adjustments in stock returns (i.e., creating news) and reductions
in market volatility (i.e., reducing noise). Whereas this mainly relates
to communication of readily processed and finalized data products,
such as on the higher levels of internal communication, it obviously
is a comparatively more challenging task due to the large heterogeneity
in the audience. 

A direct example of such communication is Financial Stability Reports,
which indeed can, and already to some extent do, make use of visual
representations to communicate the state of financial stability. Relating
to the study of the Riksbank's financial stability work by \citet{Allenetal2004},
the last exemplified recommendation highlighted the importance of
providing the underlying data and making charts easily downloadable.
Beyond transparent data and visuals, the discussion also stresses
overall guidelines in presenting data in a graphical fashion and overall
communication through visuals. Most importantly, the authors highlight
that Financial Stability Reports do and should contain a wealth of
data and indicators, the data ought to be presented through graphical
means and the graphical means ought to be presented in an easily accessible
and understandable fashion, which is not ''too busy''. This somewhat
paradoxical conclusion may also be seen as a call for interaction
techniques, with which large amounts of data can be explored but filtered
in ways that support understanding.

The most common representation of multidimensional data, yet not interactive,
is based upon work by \ac{IMF} staff on the \ac{GFSM} \citep{Dattels2010},
which has sought to disentangle the sources of risks by a mapping
of six composite indices with a radar-chart visualization. The aim
of the GFSM coincide well with those of external risk communication:
\emph{''a summary tool for communicating changes in the risks and
conditions {[}...{]} in a graphical manner {[}...{]} to improve the
understanding of risks and conditions {[}...{]} and ultimately to
warn policymakers and market participants about the risks of inaction.''}
Relating to the use of judgment, the GFSM not only leaves it to the
eyes of the beholder, but goes even further by making use of judgment
and technical adjustment to the data prior to visualization. Again,
one key task is to achieve transparency and accountability, but obviously
this time at an external level. To relate to Information Visualization
and Visual Analytics, the task of external communication clearly focuses
on spreading rather than creating knowledge, and is hence better aligned
with the former approach to visualization.

\subsection{Macroprudential data\label{sub:Macroprudential-data}}

To arrive at the data used for macroprudential oversight, we need
to recall that Section \ref{sub:Analytical-tools-for-macropru} related
analytical tools in macroprudential oversight to risk identification
and assessment. Along these lines, \citet{Borio2009} illustrates
how a macroprudential approach to fi{}nancial regulation and supervision
is best thought of as consisting of two respective dimensions: the
time and cross-sectional dimensions. First, the time dimension refers
to how systemic risk evolves over time and relates to the procyclicality
of the fi{}nancial system. Second, the cross-sectional dimension refers
to how systemic risk is distributed in the fi{}nancial system at a
given point in time and relates to common exposures across fi{}nancial
intermediaries and the systemic risk contribution of each institution.
This relates data needs to entities and time. Moreover, early-warning
exercises most often also make use of a wide range of indicators,
measuring various dimensions of risks, vulnerabilities and imbalances.
In particular, macroprudential data can be related to three different
categories of indicators: \emph{i}) macroeconomic data, \emph{ii})
banking system data, and \emph{iii}) market-based data.

Generally, the key three sources of macroprudential data measure the
behavior of three low-level entities: households, firms and assets.
By grouping data for the entities, we may produce data on various
levels of aggregation. While firm-level data may be of interest in
the case of \acp{SIFI}, data for macroprudential analysis refers
oftentimes to high-level aggregations of three kinds (see, e.g., \citet{Woolford2001}):
macroeconomic, banking system, and financial market behavior. Accordingly,
the low-level entities may be aggregated as follows: from data on
individual households' actions to the macroeconomic, from data on
banks to the banking system, and from data on individual assets to
the financial market. For instance, an entity could be a country,
which would be described by country-level aggregates of macroeconomic,
banking system, and financial market behavior. Despite the importance
of the banking sector, sectoral aggregation may likewise be defined
in broader terms (e.g. financial intermediaries in general) or some
other type of financial intermediaries (e.g., insurers or shadow banks).
It is still worth to note that a system-wide approach does not always
necessitate aggregation, as a system may also be analyzed from the
viewpoint of its more granular constituents, such as characteristics
of a network of entities and the overall emergence of system-wide
patterns from micro-level data. For instance, \citet{Korhonen2013}
links the importance of micro-level data in macroprudential analysis
to a number of possibilities, such as flexibility and information
to determine appropriate subsectors, timelier pre-assessment of impacts,
more granular composition of different exposures and different scopes
of consolidation based upon the same data.

\begin{figure}
\begin{centering}
\includegraphics[width=0.9\columnwidth]{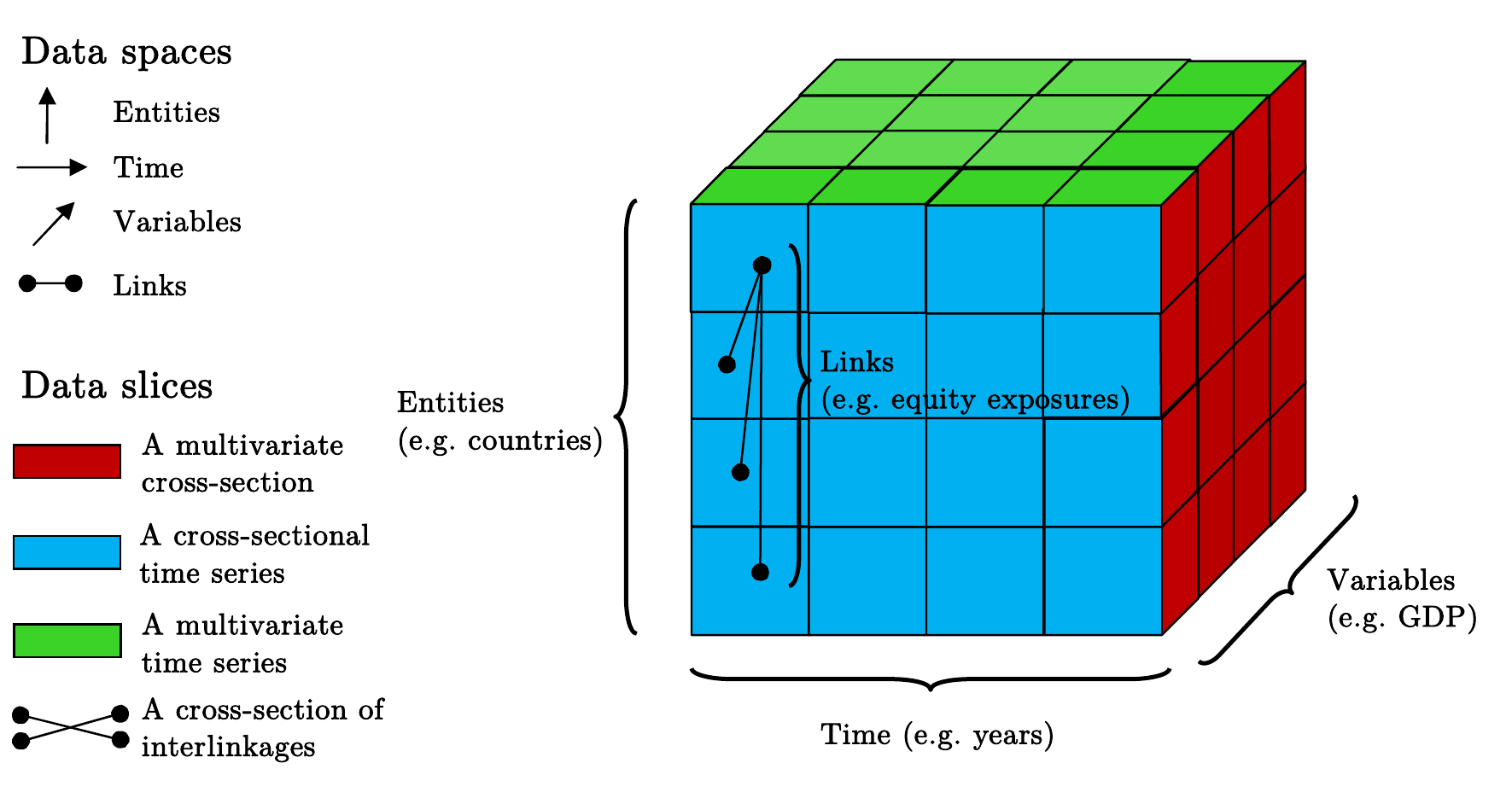}
\par\end{centering}

\noindent \textbf{\scriptsize Notes}{\scriptsize : The figure represents
the macroprudential data cube. It represents four spaces: entities
(e.g., country), time (e.g., year), variables (e.g., \ac{GDP}), and
links (e.g., debt and equity exposures). Likewise, it illustrates
four data slices: a multivariate cross section (}\textcolor{red}{\scriptsize red
side}{\scriptsize ), a cross section of time series (}\textcolor{blue}{\scriptsize blue
side}{\scriptsize ), a multivariate time series (}\textcolor{green}{\scriptsize green
side}{\scriptsize ), and a cross section of interlinkage matrices
(}\textbf{\scriptsize black edges}{\scriptsize ).}{\scriptsize \par}

\centering{}\caption{\label{Fig:financialdatacube} A macroprudential data cube.}
\end{figure}

Though we have concluded that we have three dimensions in data, entities,
time and indicators, the discussion thus far has provided little structure
on the form and complexity of the data. Independent of the aggregation
level, macroprudential oversight is most commonly utilizing structured
data that come from a so-called macroprudential data cube (henceforth
data cube). Yet, rather than three, the data cube in Figure \ref{Fig:financialdatacube}
is described by four dimensions: (\emph{i}) entities (e.g., countries);
(\emph{ii}) time (e.g., years); (\emph{iii}) indicators (e.g., credit-to-GDP
gap); (\emph{iv}) links (e.g., debt and equity exposures). Each cell
is hence defined by a specific entity, a specific time unit and a
specific variable, as well as its specific set of interlinkages. The
value for each cell is the value for that particular variable and
the related vector of links. Yet, this representation specifies little
about the size of the dataset. Beyond the hazy notion of 'big data',
this gives at hand a common setting with large-volume (cf. entities),
high-dimensional (cf. indicators) and high-frequency (cf. time) data,
where overall size of data is mostly limited by the detail level at
which each dimension of the data cube is explored. Hence, for the
three more standard dimensions of the cube, a big data problem may
arise from increases in size in any of the dimensions. Likewise, the
size of the fourth dimension largely follows entities, time and variables,
where entities refer to the number of nodes, time to time-varying
networks and variables to multi-layer networks.

Following the four dimensions, the data cube can be described according
to four types of slices. First, a multivariate cross section (\textcolor{red}{red
side}) provides a view of multiple entities described by multiple
variables at one point in time. Second, a cross section of time series
(\textcolor{blue}{blue side}) is a univariate view of multiple entities
over time. Third, a multivariate time series (\textcolor{green}{green
side}) provides a view of multiple variables over time for one entity.
Finally, the fourth view is a cross section of interlinkage matrices
(\textbf{black edges}) that represent links between multivariate entities
at one point in time. While links oftentimes refer to direct linkages
and exposures between entities, it is also common to estimate links
from interdependence in the variable dimension (e.g., stock returns).
By means of a simple example of a macroprudential dataset in the data
cube representation, the four dimensions could be defined as follows:
countries as entities, quarterly frequency as time, indicators of
various sources of risk and vulnerability as variables, and equity
and debt exposures between economies as links.

This provides a starting point to data visualization, as it ought
to be viewed from the viewpoint of the underlying data.

\subsection{Macroprudential visualization: A review}

This section provides first a brief overview of used visualization
tools for the above discussed three types of analytical tools: \emph{i})
early-warning models\emph{,} \emph{ii}) macro stress-testing models,\emph{
}and \emph{iii}) contagion and spillover models\emph{. }Whereas the
first type deals with the time dimension (i.e., risk identification),
the second and third types deal with the cross-sectional dimension
(i.e., risk assessment). Then, we discuss a categorization of visualization
methods based upon the needs for macroprudential oversight.

First, standard \emph{early-warning indicators and models} may be
complemented by the use of visualization tools for amplifying cognition.
Due to the complexity of financial systems, a large number of indicators
are often required to accurately assess the underlying risks and vulnerabilities,
and these are oftentimes compressed into a single vulnerability measure
of an early-warning model. As with statistical tables, standard two-
and three-dimensional visualizations have, of course, their limitations
for high dimensions, not to mention the challenge of including a temporal
or cross-sectional dimension or assessing multiple countries over
time. In particular, capturing the time dimension of systemic risk
is a key aim of early-warning models (e.g., \citealp{Borio2009})
Although composite indices of leading indicators and predicted probabilities
of early-warning models enable comparison across countries and over
time, these indices fall short in describing the numerous sources
of distress.

Some recent approaches make use of techniques for multidimensional
visualization to assess sources of risk and vulnerability. To start
with the \ac{GFSM}, it falls short in reducing dimensionality of
the problem, as similarity comparisons of high-dimensional observations
is left to be performed by the human eye. In addition, familiar limitations
of radar charts are, for example, the facts that area does not scale
one-to-one with increases in variables and that the area itself depends
on the order of dimensions. This can be illustrated by means of an
example, whe\textcolor{black}{re two countries hav}e both an equal
amount of aggregated risk in three subdimensions, but one has these
as neighboring axes and the other a risk in every second axis. In
this case, the former has a significantly (or infinitely) different
size but the same aggregate risks (i.e., mean value). Indeed, the
\ac{GFSM} comes with the following caveat: \emph{\textquotedblleft{}given
the degree of ambiguity and arbitrariness of this exercise the results
should be viewed merely illustrative\textquotedblright{}}.%
\footnote{The use of adjustment based on market and domain intelligence, especially
during crises, and the absence of a systematic evaluation gives neither
a transparent data-driven measure of financial stress nor an objective
anticipation of the \ac{GFSM}\textquoteright{}s future precision.
The authors state that the definitions of starting and ending dates
of the assessed crisis episodes are somewhat arbitrary. The assessed
crisis episodes are also subjectively chosen. Introduction of judgment
based upon market intelligence and technical adjustments are motivated
when the \ac{GFSM} is \emph{\textquotedblleft{}unable to fully account
for extreme events surpassing historical experience\textquotedblright{}},
which is indeed an obstacle for empirical models, but also a factor
of uncertainty in terms of future performance since nothing assures
manual detection of vulnerabilities, risks and triggers.%
} 

Mapping techniques with the aim of data reduction and dimension reduction
have also been used to represent these complex data. In terms of \ac{FCM}
clustering, a combination of clustering models and the reasoning of
fuzzy logic have been introduced to the early-warning literature by
finding risky clusters and treating relationships in data structures
as true or false to a certain degree \citep{ISAF:ISAF317}. Beyond
signaling a crisis in a timely manner, this type of analysis has the
benefit of signaling the type and degree of various sorts of financial
imbalances (in terms of memberships to clusters). In an exploratory
study, \citet{ArciniegasRuedaA09} found, with the help of the \ac{SOM},
strong associations between speculative attacks\textquoteright{} real
effects and 28 indicators, yet did neither focus on visualizing individual
data nor on early-warning performance.

Turning to SOM based papers focusing on visualization, \textcolor{black}{\citet{Sarlin2010Currency}
presents the first exploratory study of the \ac{SOM} as a display
of crisis indicators with a focus on the Asian currency crises in
1997--1998. \citet{ISAF:ISAF321} extend the work by using the \ac{SOM}
as an early-warning model, including an evaluation in terms of predictive
performance, and with a larger sample of indicators. In \citet{Sarlin2011Debt},
the \ac{SOM} is applied to a wide range of indicators of sovereign
default. Further, }\citet{SarlinPeltonen2013} create the \ac{SOFSM}
that lays out a more general framework of data and dimension reduction
for mapping the state of financial stability, and visualizing potential
sources of systemic risks. As an early-warning model, the \ac{SOFSM}
is shown to perform on par with a statistical benchmark model and
to correctly call the crises that started in 2007 in the \ac{US}
and the euro area.\textcolor{black}{{} All of these works highlight
the usefulness of the \ac{SOM} for the task.}

Second, \emph{macro stress-testing models}, to the best of my knowledge,
make no use of advanced visualization techniques for representing
the results of the tests, including the processing of data at the
input, interim and output stage. Visualization seldom goes beyond
a framework or schematic structure for the designed transmission mechanisms
in the model and plots of loss distributions in various formats. Obviously,
standard visualization techniques from graph theory may be used in
representing networks, if such are used in the models. For instance,
the macro stress-testing model by \citet{Bossetal2006}, which integrates
satellite models of credit and market risk with a network model for
evaluating default probabilities of banks, enable one to make use
of concepts from graph theory in visualizing the network structure.
Network visualizations are, however, more common in contagion models.

As said, the third group of \emph{contagion and spillover models}
commonly make use of concepts from graph or network theory to visualize
the structure of linkages in the models (see, e.g., \citet{Estrada2011}).
This provides means to represent entities as nodes (or vertices) and
their links as edges (or arcs). The combination of nodes and edges
provide all constituents for a network, where the edges may be directed
\emph{vs}. undirected and weighted \emph{vs}. unweighted. However,
rather than a visualization, a network is a data structure. The interpretability
of networks has been enhanced by the means of various methods. For
instance, positioning algorithms, such us force-directed layout methods,
are commonly used for locating nodes with similar edges close to each
other, as well as ring and chord layouts for more standardized positioning.
Yet, the so-called hairball visualization, where nodes and edges are
so large in number that they challenge the resolution of computer
displays, not to mention interpretation, is not a rare representation
of complex financial networks (see, e.g., \citet{BechAtalaya2010}).
By including dynamics, there exists also work on visualizing how shocks
cascade in networks, which directly relates to the contagion literature
(e.g., \citealp{vonLandesberger2014}). It is worth noting that recent
advances in financial network analysis, particularly software, hold
promise in bringing aesthetics and the ease of use of visualizations
into the financial domain. An additional essential feature, not the
least to deal with hairballs, is the use of interaction techniques
with visualizations.

\subsection{The need for analytical and interactive visualizations}

This section has so far defined the task of visualization in risk
communication, the structure and properties of macroprudential data
and visualizations used to date to represent systemic risk. Beyond
the indications of what has been done so far in the state-of-the-art
overview, the notions of risk communication and the underlying data
highlight the need for two features: \emph{i}) interactive and \emph{ii})
analytical visualizations. While analytical techniques enable the
visualization of big data, interactivity answers the needs set by
communication as it enables extracting large amounts of information
through the interaction with a graphic.

The need for \emph{analytical techniques} refers to the complexity
of data used in systemic risk measurement. Visual analytics refers
often to the coupling of visual interfaces to analytics, which supports
in building, calibrating and understanding models (e.g., early-warning
models). Yet, the notion of an analytical technique for visualization
differs by rather using analytics for reducing the complexity of data,
with the ultimate aim of visualizing underlying data structures. These
techniques provide means for drilling down into the data cube. For
instance, mapping techniques provide through dimension reduction a
projection of high-dimensional data into two dimensions, whereas clustering
methods enable reducing the volume of data into fewer groups (or mean
profiles). Likewise, the analysis of time might also be supported
by the use of analytical techniques, as compressing the temporal dimension
would enable representations of only relevant points in time.

The coupling of visual interfaces with \emph{interaction techniques}
goes to the core of information visualization and visual analytics.
One key task of macroprudential supervisors has been to publish risk
dashboards, such as that of the ESRB, yet none of these have been
truly interactive. For instance, the sixth issue of the dashboard
was a static 30 page document, in addition to 11 pages of annexes.
While the initiative has merit in that it promotes transparency and
knowledge among the general public, it is not clear why the dashboard
is lacking interactivity and true data sharing in an age when the
single most used distribution channel is digital format.%
\footnote{It is still worth to note that ESRB's risk dashboard has been paired
with the ECB Statistical Data Warehouse. Yet, data for far from all
indicators is shared in the datawarehouse.%
} Even though one could argue that formal reports, such as Fed's Annual
Report to Congress, require static visuals to support transparency,
accountability and record-keeping, the property of exporting a static
screenshot is a common property of interactive interfaces. To the
other extreme, if accountability and common knowledge is of utmost
importance and limits the design of reports, then one could ask whether
any visuals should be utilized. Another example is when a systemic
risk model (e.g., early-warning model) has been built and calibrated.
It would be an intuitive first step to circulate it (at least internally)
with possibilities to explore all examples of potential interest,
rather than only being presented with cases selected by its authors.
While this could function as a type of review or scrutiny at an early
development stage, this would also support transparency, and thus
possibly credibility, of internally developed approaches to identify
and assess systemic risk. Further, relating to the previous topic,
a natural extension to analytical visualization techniques would obviously
be means to interact with them.

\section{Visualization applications and the VisRisk platform}

This section moves from previous abstractions to concrete applications.
With the two features pinpointed in the previous section and the macroprudential
data cube in mind, this section provides a range of examples both
relating to analytical and interactive visualizations in macroprudential
oversight. First, we make use of analytical techniques for data and
dimension reduction to explore high-dimensional systemic risk indicators
and time-varying networks of linkages. Second, we add interactivity
to not only dashboards of standard risk indicators and early-warning
models, but also to the analytical applications. Hence, we illustrate
applications of three analytical visualizations and five interactive
web-based visualizations to systemic risk indicators and models. From
the viewpoint of the data cube in Figure \ref{Fig:financialdatacube},
we provide visual means to explore all four dimensions. 

Beyond the applications herein, it is worth remembering that the ultimate
aim of the paper is to provide a platform or basis for the use of
visualization techniques, especially those including analytical and
interactive features, in macroprudential oversight in general and
risk communication in particular. Hence, we end this section by presenting
the VisRisk platform that enables and is open to the visualization
of any data from the macroprudential data cube.

\subsection{Analytical visualizations\label{sub:Analytical-visualizations}}

This subsection presents three cases where data and dimension reduction
techniques are used for representing large-volume and high-dimensional
data in simpler formats. Further, we also present an application of
a force-directed layout algorithm to the visualization of network
data. Even though we address systemic risk, it is a deliberate choice
not to make use of data that would be overwhelmingly extensive along
any of the dimensions of the data cube. This is mainly due to the
illustrative nature of the examples, as the main function of this
section is to demonstrate the usefulness of analytical and interactive
visualizations in macroprudential oversight. Nevertheless, the applications
make use of real-world data targeted at systemic risk.

The problem setting described in Section \ref{sub:Macroprudential-data},
particularly the combination of large-volume and high-dimensional
data, has been addressed in machine learning, a subfield of computer
science. One common approach to reducing large-volume data (i.e.,
many high-dimensional observations) is to represent a set of points
$P$ by a smaller, but representative, set of equidimensional points
$A$. These representative points constitute partitions or clusters
and may be called reference vectors, mean profiles or cluster centroids,
mostly with an aim to approximate the probability density functions
of data. Data reduction is a necessity, and thus common practice,
in application areas like speech recognition, computational linguistics,
and computational genomics, and is addressed by means of clustering
algorithms, or so-called data reduction techniques (see, e.g., \citet{Jain2010}).
While both $P$ and $A$ still live in the high-dimensional space
$\mathbb{R}^{n}$, machine learning also provides approaches for reducing
the dimensionality of data. Dimension reduction provides low-dimensional
overviews of similarity relations in high-dimensional data, where
data are represented in two dimensions such that similar high-dimensional
data are nearby and dissimilar distant. The task of projecting  observations
$p_{j}\in\mathbb{R}^{n}$ to an equisized set of $l_{j}\in\mathbb{R}^{2}$
goes also by the name of manifold learning, embedding and mapping
(see, e.g., \citet{LeeVerleysen2007}), but can also be considered
to cover force-directed layout algorithms for positioning nodes of
networks. A recent focus has been to also involve the third dimension
-- time. This refers to the task of visual dynamic clustering, where
\emph{clustering }refers to reducing data, \emph{visual }refers to
reducing dimensionality and \emph{dynamic} refers to changes in clusters
over time. Hence, it provides means for visualizing how cross-sectional
structures (i.e., clusters) evolve over time.

\subsubsection*{Financial Stability Map}

\begin{figure}
\begin{centering}
\includegraphics[width=0.9\columnwidth]{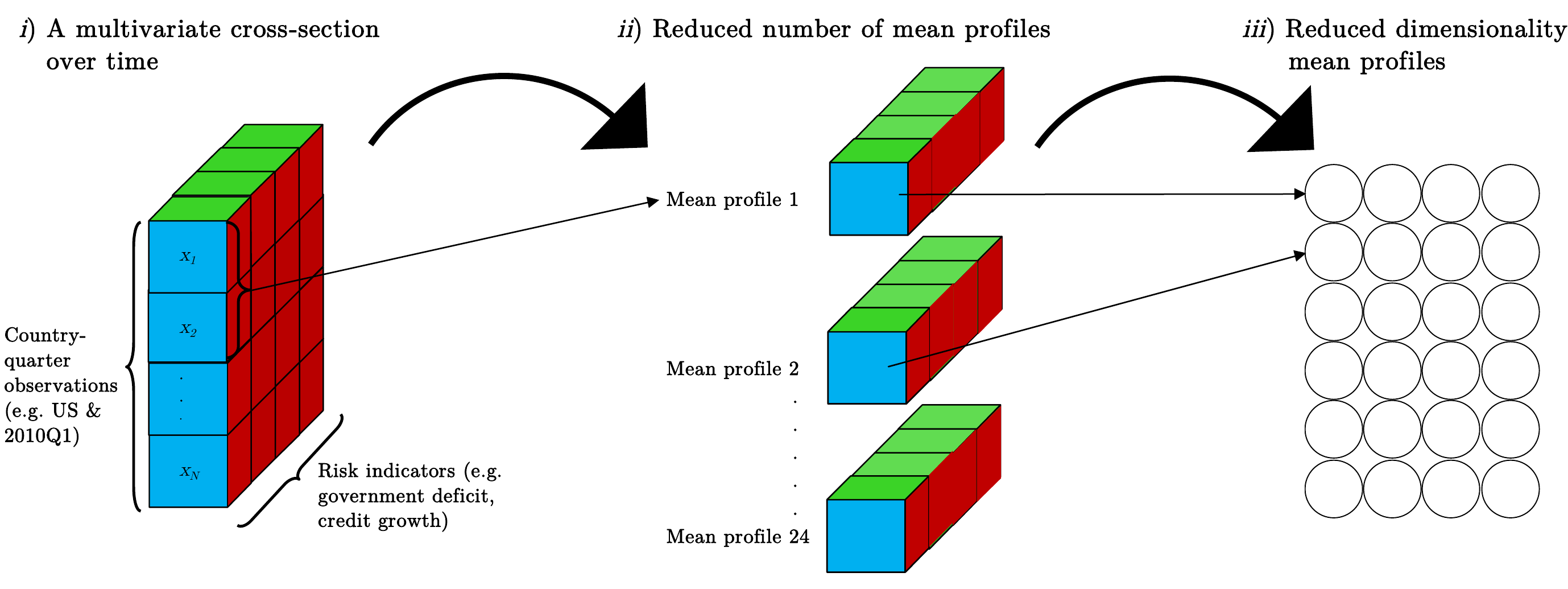}
\par\end{centering}

\noindent \textbf{\scriptsize Notes}{\scriptsize : The figure uses
the data structure presented in the data cube to describe how the
SOM-based data and dimension reduction is performed. By pooling the
cross-sectional and temporal dimensions, the }\textcolor{red}{\scriptsize red
side}{\scriptsize{} represents a multivariate cross section over time,
in which the }\textcolor{blue}{\scriptsize blue side}{\scriptsize{}
represents time-entity observations and the }\textcolor{green}{\scriptsize green
side}{\scriptsize{} common indicators. }{\scriptsize \par}

\centering{}\caption{\label{Fig:SOM processing} Creating the Financial Stability Map.}
\end{figure}

This part describes the application of the (Self-Organizing) Financial
Stability Map (FSM), as originally described in \citet{SarlinPeltonen2013}.
The underlying data come from the three standard dimensions of the
data cube: quarterly observations for a global set of 28 economies
from 1990--2011 and 14 macro-financial indicators. The FSM is based
upon the Self-Organizing Map (SOM), which has two aims: \emph{i})
to reduce large amounts of high-dimensional data to fewer mean profiles,
and \emph{ii}) to provide a low-dimensional representation of the
high-dimensional mean profiles. As described in Figure \ref{Fig:SOM processing},
using a pooled panel data (i.e., time is not taken into account) in
step \emph{i}), the FSM follows two subsequent steps: \emph{ii})\emph{
}to\emph{ }group high-dimensional observations to a smaller set of
equidimensional mean profiles based upon similarity, and\emph{ iii})
to project the mean profiles to an equisized but low-dimensional grid
such that similar profiles are located close by.

In this application, the motivation for using the SOM for mapping
financial stability over alternative techniques relates mainly to
the following properties (for a further discussion see \citet{SarlinIV2013}):
\emph{i}) its simultaneous clustering and projection capabilities,
\emph{ii}) the pre-defined grid structure for linking visualizations,
\emph{iii}) computational efficiency, and \emph{iv}) flexibility for
missing data. For a description of the technical details, readers
are referred to Appendix B.1 and \citet{Kohonen1982,Kohonen2001}.

\begin{figure}
\begin{centering}
\includegraphics[width=0.5\columnwidth]{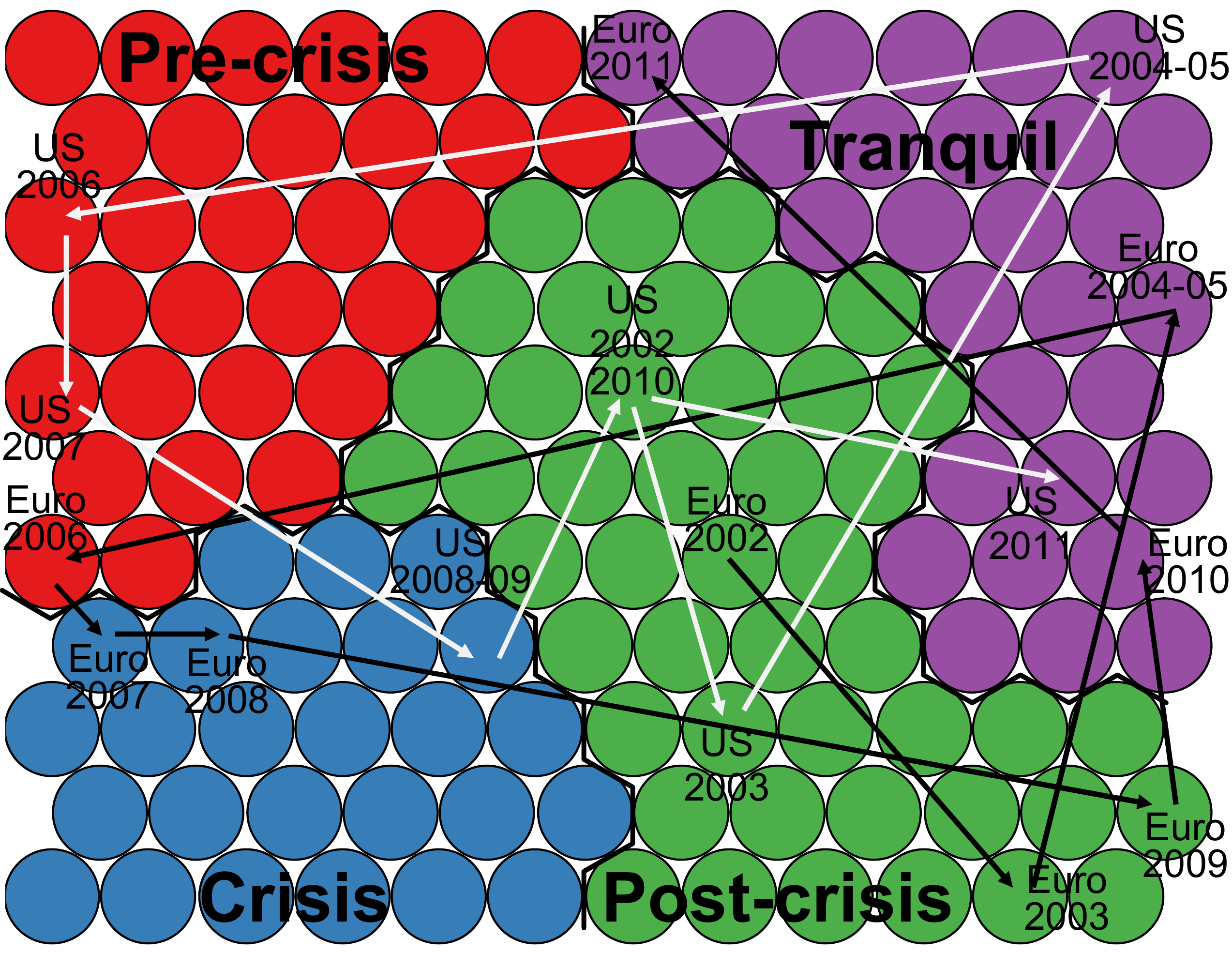}\includegraphics[width=0.5\columnwidth]{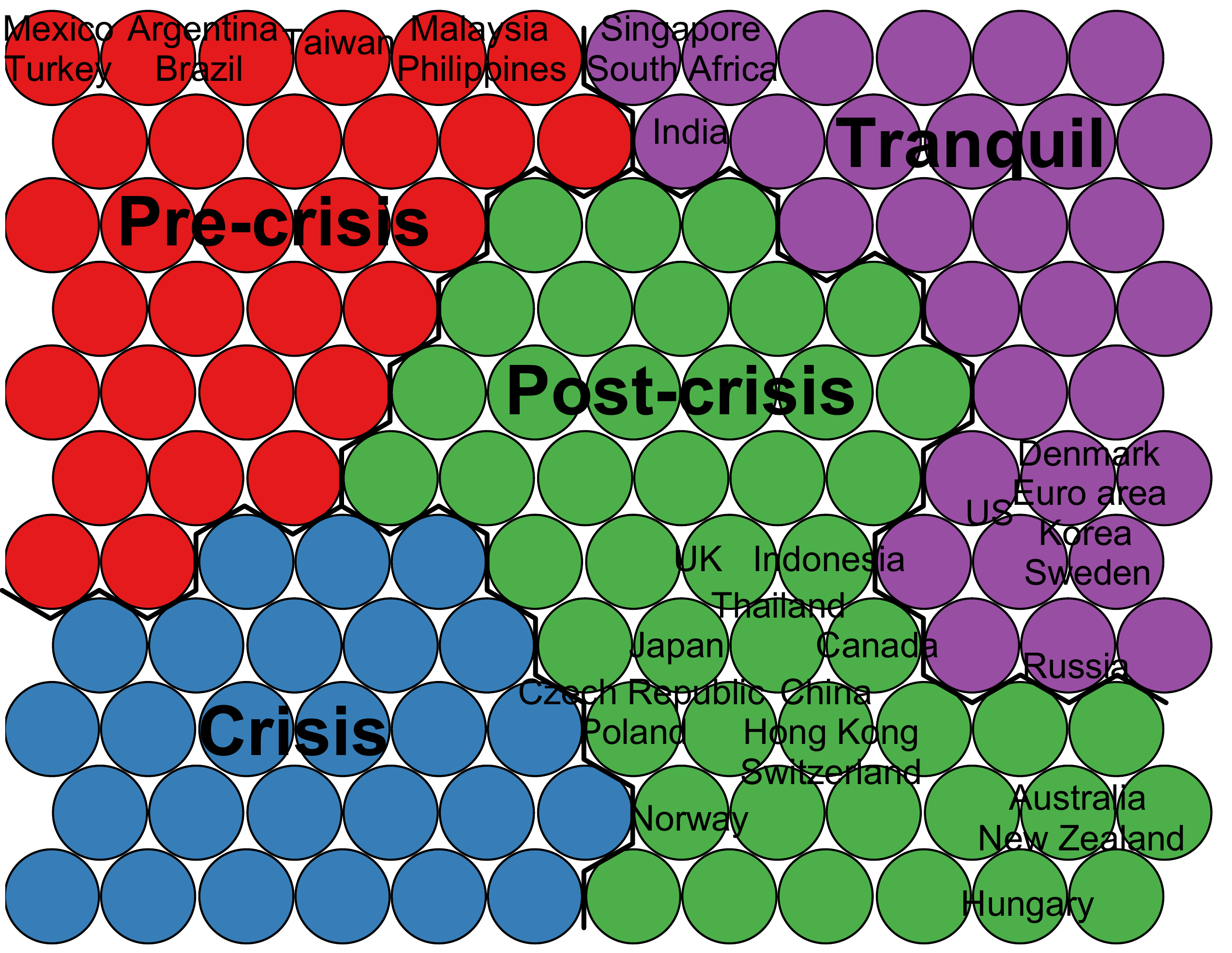}
\par\end{centering}

\textbf{\scriptsize Notes}{\scriptsize : The figure displays the two-dimensional
FSM that represents a high-dimensional financial stability space.
The lines that separate the map into four parts are based on the distribution
of the four underlying financial stability states. Data points are
mapped onto the grid by projecting them to their \acp{BMU} using
only macro-financial indicators. Consecutive time-series data are
linked with arrows. In the left figure, the data for both the \ac{US}
and the euro area represent the first quarters of 2002--2011 as well
as the second quarter of 2011. In the right figure, the data for all
economies represent the third quarter of 2010.}{\scriptsize \par}

\centering{}\caption{\label{Fig:SOM applications} Evolution in the US and euro area and
a cross section on the FSM.}
\end{figure}

The procedure described in Figure \ref{Fig:SOM processing} creates
the FSM, and provides hence a low-dimensional basis or display with
two tasks: \emph{i}) to function as a display for visualizing individual
data concerning entities and their time series, and \emph{ii}) to
use the display as a basis to which additional information can be
linked. As the nodes of the grid are high-dimensional mean profiles,
the high-dimensional observations can be located with their most correct
position using any similarity measure. In Figure \ref{Fig:SOM applications},
we illustrate the evolution of macro-financial conditions (14 indicators)
for the United States and the euro area (2002--2011, first quarter),
and a cross section of macro-financial conditions in key advanced
and emerging market economies in 2010Q3. Beyond these temporal and
cross-sectional applications, the FSM could be paired other approaches
and applications, such as aggregated data (e.g., world conditions),
scenario analysis by introducing shocks to current conditions, assessing
linkages by connecting linked economies with edges, etc. (see \citet{Sarlinthesis2013}
for further examples).

Yet, this map does not provide means for understanding how the cross
sections change over time.

\subsubsection*{Financial Stability Map over Time}

\begin{figure}
\begin{centering}
\includegraphics[width=0.85\columnwidth]{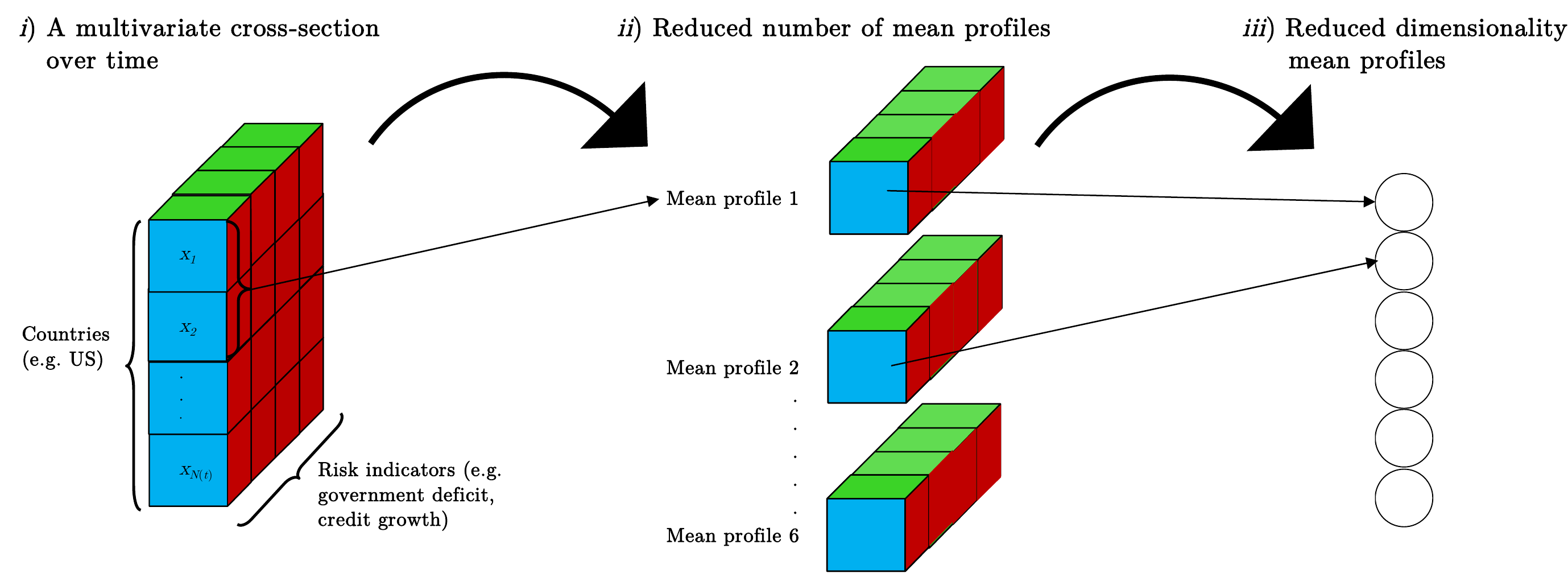}
\par\end{centering}

\begin{centering}
\includegraphics[width=0.65\columnwidth]{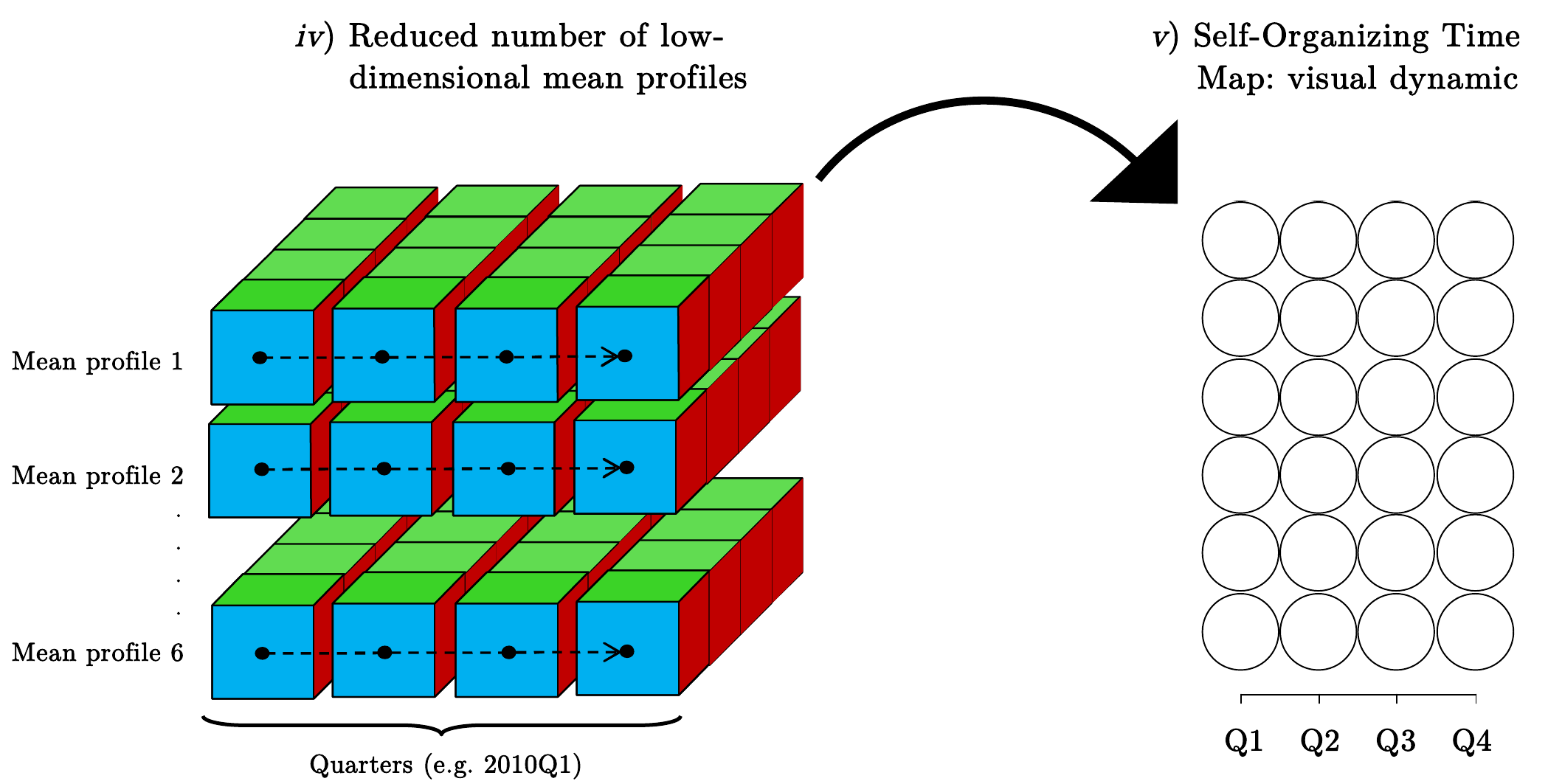}
\par\end{centering}

\noindent \textbf{\scriptsize Notes}{\scriptsize : The figure uses
the data structure presented in the data cube to describe how the
SOTM-based data and dimension reduction is performed. Instead of pooling
data, the SOTM applies one-dimensional SOMs to individual high-dimensional
(}\textcolor{red}{\scriptsize red side}{\scriptsize ) cross sections
(}\textcolor{blue}{\scriptsize blue side}{\scriptsize ). As this is
performed over time (}\textcolor{green}{\scriptsize green side}{\scriptsize ),
in conjunction with techniques for initialization and a short-term
memory, the orientation of the one-dimensional SOMs is preserved.}{\scriptsize \par}

\centering{}\caption{\label{Fig:SOTM processing} Creating the Financial Stability Map
over Time.}
\end{figure}

To understand how cross sections are evolving, we need to tap into
approaches for visual dynamic clustering. The recently introduced
Self-Organizing Time Map (SOTM) \citep{Sarlin2012} is unique in that
it provides means for visualizing how cross-sectional structures (i.e.,
clusters) evolve over time. Although the only difference to the above
approach is the addition of a time dimension to the mapping, it provides
an approach that truly represents the three dimensions of the data
cube. Hence, it addresses large-volume data by reducing the number
of entities to clusters and high-dimensional data by reducing dimensionality
the clusters, as well as represents changes in clusters over time.
Based upon the SOTM, this part describes the application of the Financial
Stability Map over Time (FSM-t), as originally described in \citet{SarlinPRL2013}.
In contrast to the above application, we do not pool the panel data,
but rather follow the first three steps as described in the upper
part of Figure \ref{Fig:SOTM processing}: \emph{i}) starting from
a multivariate cross section, \emph{ii}) high-dimensional observations
are grouped to a smaller set of equidimensional mean profiles and\emph{
iii}) the mean profiles projected to a one-dimensional grid such that
similar profiles are close by. Next, as this concerns only an individual
point in time, steps \emph{iv}) and \emph{v}) in the lower part of
Figure \ref{Fig:SOTM processing} describe performing the three former
steps on all time points, including a number of initialization techniques
to preserve orientation, in order to create a two-dimensional grid
representing the spaces of both time and data. For technical details
on the SOTM, readers are referred to Appendix B.2 and \citet{Sarlin2012}.

\begin{figure}
\begin{centering}
\includegraphics[width=0.75\columnwidth]{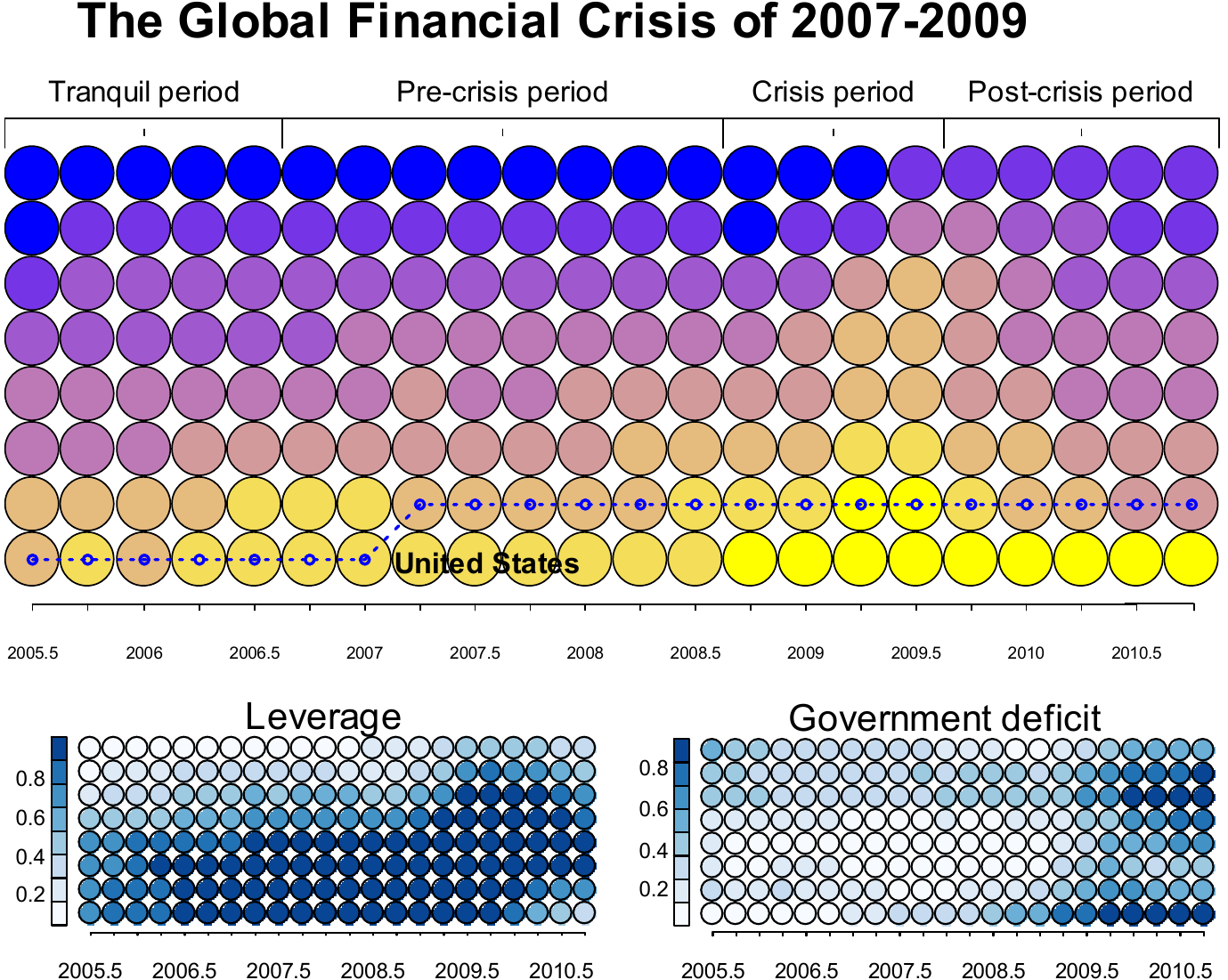}
\par\end{centering}

\textbf{\scriptsize Notes}{\scriptsize : The figure represents a \ac{SOTM}
of the global financial crisis, where the cluster coloring shows changes
in multivariate cluster structures. Labels above the figure define
the classes in data, i.e., the stages of the financial stability cycle,
and the trajectory on the \ac{SOTM} represents the evolution of macro-financial
conditions in the \ac{US} .}{\scriptsize \par}

\centering{}\caption{\label{Fig:SOTM applications} A \ac{SOTM} of the global financial
crisis.}
\end{figure}

The procedure described in Figure \ref{Fig:SOTM processing} creates
the FSM-t. Although the FSM-t can also function as a display for visualizing
individual data concerning entities and their evolution, the key task
of it is to focus on how the high-dimensional structures are evolving
over time. In Figure \ref{Fig:SOTM applications}, the upper figure
illustrates the evolution of macro-financial conditions (i.e., 14
indicators) in the cross section from 2005Q3--2010Q4. In terms of
all 14 indicators, similarity in the mean profiles is represented
by similarity in color (from blue to yellow). Further, as the nodes
of the grid are high-dimensional mean profiles, one particularly interesting
view is to explore how individual variables have evolved in the cross
section over time. The lower part of Figure \ref{Fig:SOTM applications}
shows how leverage is increasing over time, even after the crisis
of 2007--2008, and the increases in government deficits after the
first wave of the crisis. The FSM-t can also be created and exploited
in various different ways, such as focusing on cross-sectional changes
before, during and after crises (e.g., $t-8,t-7,...,t+8$) and by
projecting individual economies on top of it.

Now, we have tackled the three standard dimensions of the data cube,
but what about the linkages?

\subsubsection*{Bank Interrelation Map}

Like high-dimensional risk indicators, networks constitute an inherently
complex source of information. Network analysis, or link analysis,
can be seen as the exploration of crucial relationships and associations
between a large set of objects, also involving emergent properties,
which may not be apparent from assessing isolated data. Networks of
relationships are mostly expressed in matrix form, where the link
between entities $g$ and $l$ in a matrix $A$ is represented by
element $a_{gl}$. The matrix is of size $n^{2}$, where $n$ is the
number of entities. Matrices of directed graphs can be read in two
directions: rows $g$ of $A$ represent the relationship of $g$ to
$l$ and columns $l$ of $A$ represent the relationship of $l$ to
$g$. Each entity $g$ is thus described by its relationship to each
other entity $l$, and hence $g\in\mathbb{R}^{n}$. Yet, except for
the rare completely connected networks, dense networks comprise for
each node an edge number close to the total number of nodes, whereas
the nodes in sparse networks comprise only a low number of links,
which can also be called scale-free networks if their degree distribution
follows a power law. The latter type of networks are predominant in
the real world, where the network structure is formed through natural
processes and central hubs naturally emerge. In the vein of the above
defined task of dimension reduction, this describes the challenge
of drawing low-dimensional graphs from complex networks.

There is a range of approaches for visualizing network data. One popular
category is that of force-directed layout algorithms. They provide
means for representing network centrality and connectivity (i.e.,
dependency structures) in low dimensions, by optimizing node positions
according to some of the following criteria, among others: few crossing
edges, straight edges, long enough edges and uniform edge length for
non-weighted networks. Based upon work in \citet{RonnqvistSarlin2014},
this part describes the use of financial discussion for creating bank
interrelation maps. Most analysis of interdependencies among banks
has been based on numerical data. By contrast, this study attempts
to gain further insight into bank interconnections by tapping into
financial discussion. The approach is illustrated using a case study
on Finnish financial institutions, based on discussion in 3.9M posts
spanning 9 years in an online discussion forum belonging to a major
Finnish web portal dedicated to financial and business information.
Based upon these data, co-occurrences of bank names are turned into
a network, which can be visualized. For the purpose of data exploration,
particularly visualization, the usage of textual data holds an additional
advantage in the possibility of gaining a more qualitative understanding
of an observed interrelation through its context. 

\begin{figure}
\begin{centering}
\includegraphics[width=0.88\columnwidth]{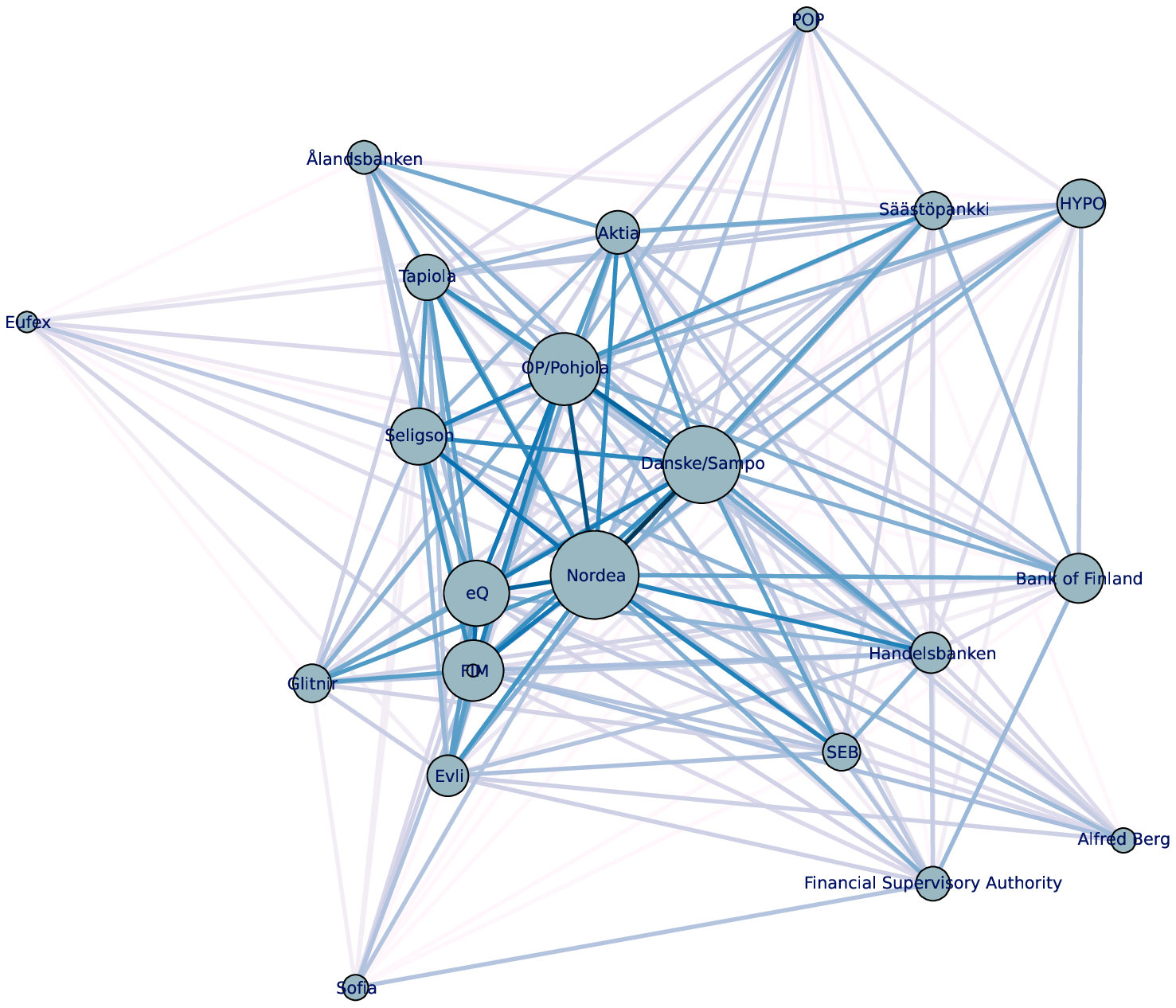}
\par\end{centering}

\textbf{\scriptsize Notes}{\scriptsize : The figure shows one solution
of a bank co-occurrences map with the Fruchterman-Reingold algorithm.
Node size is proportional to the individual bank occurrence count,
while connection darkness is logarithmically scaled to co-occurrence
count.}{\scriptsize \par}

\centering{}\caption{\label{Fig:Network banks} A bank co-occurrence network from financial
discussion.}
\end{figure}

Figure \ref{Fig:Network banks} shows a bank interrelation map based
upon a co-occurrence network from financial discussion. The network
depicts counts of bank names co-occurring in forum posts for the entire
period 2004--2012. Node size is proportional to the individual bank
occurrence count, while connection darkness is logarithmically scaled
to co-occurrence count. Nodes are positioned by the Fruchterman-Reingold
algorithm \citep{FruchtermanReingold1991}. The message of the figure
is that the largest nodes in the center of the graph are also the
most centrally connected in the network. Yet, textual data being a
rich data source, there is plenty of descriptive information to be
explored behind the linkages and nodes, something that could easily
be conveyed through interaction possibilities.

Further relating to the limits of a single, static overview is the
shortcoming of force-directed layout algorithms when processing large-volume
data. Not only are they computationally costly, but they also often
find locally optimal solutions. While there is a range of solutions
to decrease computational cost, such as multilevel \citep{Hu2006}
and parallel \citep{TikhonovaMa2008} extensions, the problem of local
optima derives from the complexity of projecting from high-dimensional
data to low-dimensional displays. One cure may, however, be the possibility
to interact with node positions, as well as other techniques improving
representation, which takes us to the following topic.

\subsection{Interactive interfaces}

This section, while still exemplifying visualizations in macroprudential
oversight, shifts the focus towards interaction with the visual interfaces.
Rather than an ending point, the visual interface or visualization
is a starting point for data exploration, to which interactivity is
an ideal support. Beyond interactive interfaces, it is worth noting,
as above discussed, that screen captures (i.e., pdf, svg and png formats)
and URL outputs (i.e., a permalink with chosen parameters) are available
to support common-information sharing.

Following Shneiderman's \citeyearpar{Shneiderman1996} visual information
seeking mantra of ''\emph{Overview first, zoom and filter, then details-on-demand}'',
the visualization provides merely a high-level overview, which should
be manipulated through interaction to zoom in on a portion of items,
eliminate uninteresting items and obtain further information about
requested items. To this end, a large share of the revealed information
descends from manipulating the medium, which not only enables better
data-driven communication of information related to risk, but also
facilitates visual presentation of big data.

\subsubsection*{Risk dashboard}

In recent years, risk dashboards have become essential tools for both
internal and external risk communication among macroprudential supervisors.
While central banks commonly have their own internal risk dashboards,
supervisory bodies like ESRB and EBA publish dashboards also available
for external assessment of prevailing risks and vulnerabilities. As
an alternative to static documents, this part introduces an interactive
data-driven document (D3) based risk dashboard of 14 systemic risk
indicators. The dashboard includes quarterly data for a global set
of 28 economies and ranges from 1990 to 2011. The dataset is based
upon that in \citet{SarlinPeltonen2013}, of which further details
are available in \citet{Sarlinthesis2013}, and which to a large extent
descends from \citet{Duca2012}.

\begin{figure}
\begin{centering}
\includegraphics[width=1\columnwidth]{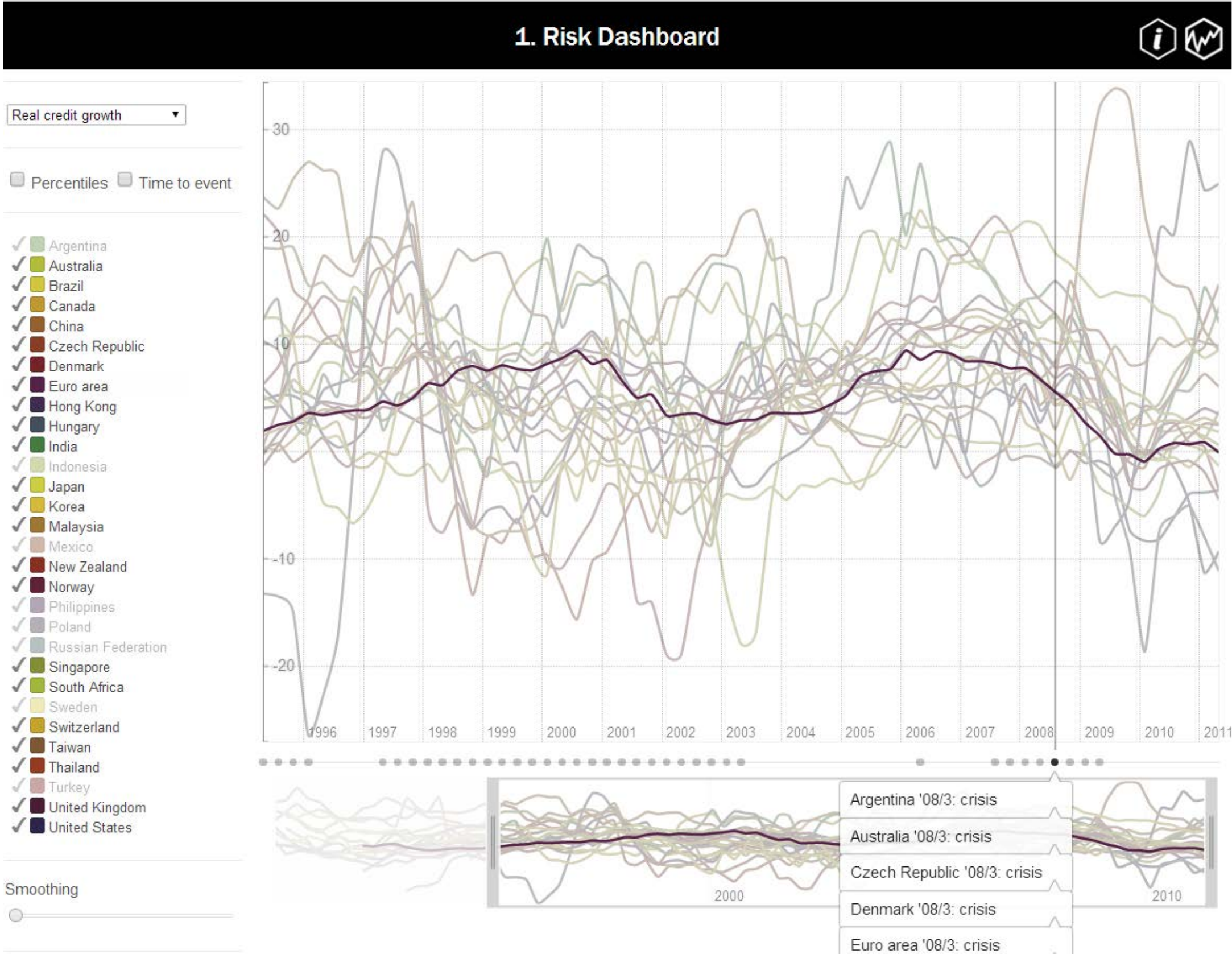}
\par\end{centering}

\textbf{\scriptsize Notes}{\scriptsize : The interactive risk dashboard
can be found here:} {\scriptsize \href{http://vis.risklab.fi/}{http://vis.risklab.fi/}.
In the screenshot, credit growth is shown for economies selected in
the left panel and hovering over the label of the euro area dims the
time series for all other economies. The time brush below the figure
is used to focus on a specific time span, whereas a drop-down list
of events is chosen to be shown from the event line between the two
displays, which also adds a vertical line to the above plot. The data
are shown in non-transformed format.}{\scriptsize \par}

\centering{}\caption{\label{Fig:Risk dashboard} An interactive risk dashboard.}
\end{figure}

The dashboard in Figure \ref{Fig:Risk dashboard} focuses on time
series of univariate indicators for a cross section. As an \emph{overview},
it presents a time-series plot for one chosen indicator and all economies,
where the indicator and its transformation can be chosen from the
drop-down menu and radio buttons, respectively. The transformation
scales the indicators to country-specific percentiles, which enables
a view of the data in relation to their own economy's distribution.
\emph{Zooming and filtering} involves highlighting individual economies
by hovering, showing only one individual economy by selection (which
highlights their events), dropping an economy from the graph by deselecting
it, and choosing a time span to be shown. \emph{More details} can
be obtained about the historical occurrence of crises in economies,
more precise information about highlighted data points (value, year
and country) and through any of the zooming or filtering options as
the entire graph (\emph{x} and \emph{y} axes) adapts to changes. Moreover,
with the same functionality, the dashboard also allows for focusing
on country-level time-to-event plots, which illustrates crisis dynamics
for all economies.

While this provides an interface to time series of a large number
of indicators, these are oftentimes combined into a single composite
indicator through various analytical techniques, which likewise would
benefit from a visual interface.

\subsubsection*{Early-warning model}

This risk dashboard of 14 systemic risk indicators provides a view
of individual indicators and their percentile transformations. These
indicators could, however, be an input to an early-warning model,
to which a visual interface would provide ample means for better understanding
the performance and characteristics of the model. Visual means also
allows for better scrutiny, which is likely to impact model credibility.

\begin{figure}
\begin{centering}
\includegraphics[width=1\columnwidth]{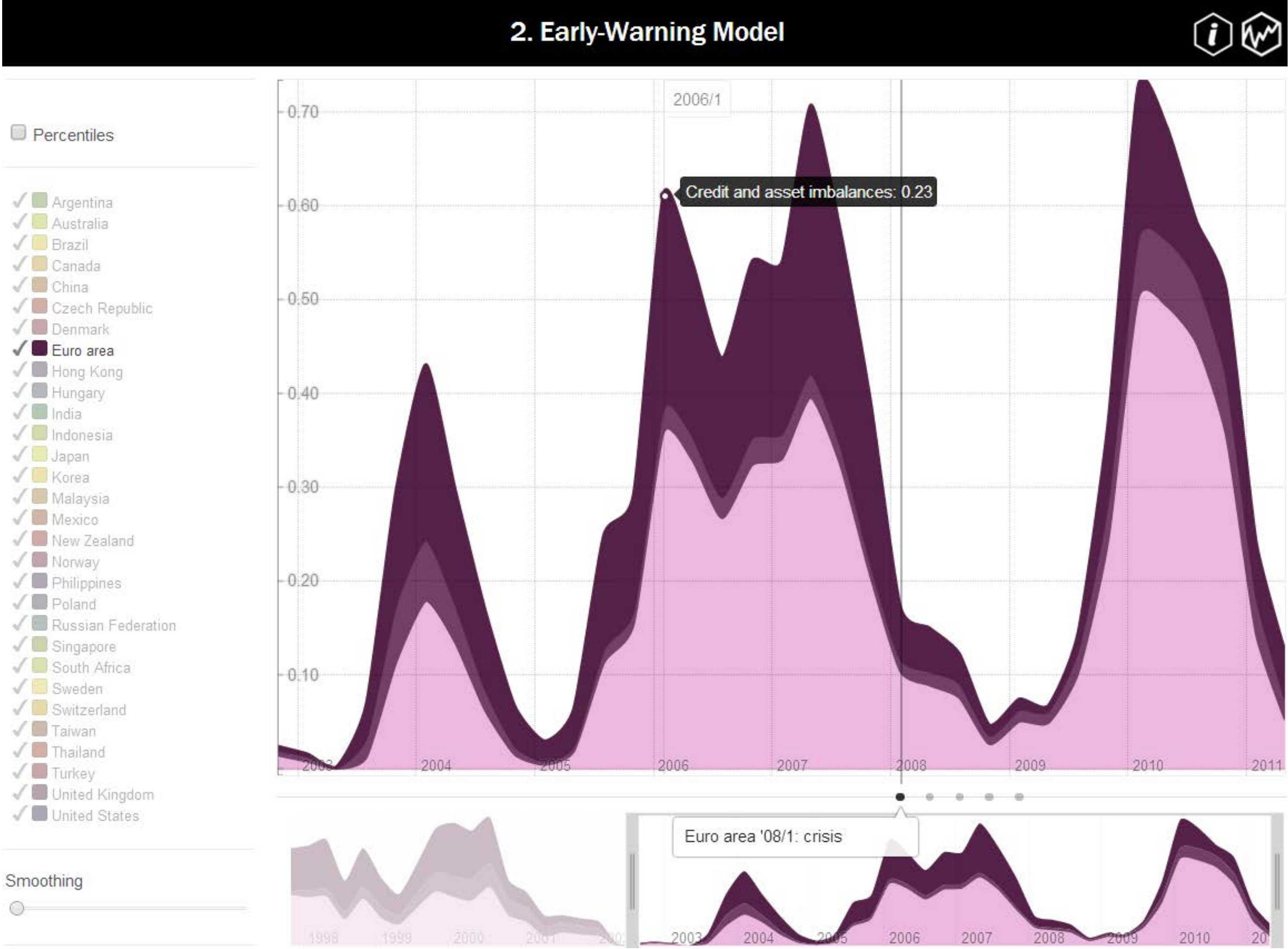}
\par\end{centering}

\textbf{\scriptsize Notes}{\scriptsize : The interactive early-warning
model can be found here: \href{http://vis.risklab.fi/}{http://vis.risklab.fi/}.
In the screenshot, estimated probabilities of systemic risk are shown
for the euro area for a specific time span, as is selected in the
left panel and the below time brush, and hovering over a specific
data point shows that the contribution, variable category and time
point are 0.23, credit and asset imbalances and 2006Q1, respectively.
The event for the euro area are chosen to be shown from the event
line above the time brush, which also adds a vertical line to the
above plot. The data are shown in non-transformed format.}{\scriptsize \par}

\centering{}\caption{\label{Fig:EWM} An interactive early-warning model.}
\end{figure}

In this part, we provide a similar interface as that for the risk
dashboard, but instead visualize the output of an early-warning model
with the previously explored early-warning indicators as an input.
The \emph{overview} illustrates how systemic risk or vulnerability
has evolved in all economies over time. While the percentile transformation
aids in interpreting the severity of the measure in its historical
distribution, we can make use of the same features for \emph{zooming
and filtering} and retrieving more \emph{details on demand} to better
understand the aggregate risk measure. As is shown in Figure \ref{Fig:EWM},
a further interaction capability is that the line graph is changed
into a stacked graph when a single economy has been chosen, which
shows the contribution of three variable groups: domestic macroeconomic
and credit and asset imbalances and global imbalances.

The importance of visualizing these types of analytical models derives
from their complexity and lack of transparency, as they are often
the result of extensive computations, within which it might not always
be easy or possible to track the quality of the output. While not
being a substitute for more formal evaluations of signaling quality
and other quantitative performance measures, this provides a means
for a more detailed and general exploration of model outputs. Moreover,
one essential task ought to be to improve transparency of analytical
projects and models within an organization, by enabling other analysts,
groups and divisions to interact with derived models, which could
function as an internal model audit.

Early-warning models do not, however, provide a more detailed view
of what is driving results and where in a financial stability cycle
an economy might be located. They give a mere probability of a crisis,
which turns our attention to the FSM and potential to interact with
it.

\subsubsection*{Financial Stability Map}

\begin{figure}
\begin{centering}
\includegraphics[width=1\columnwidth]{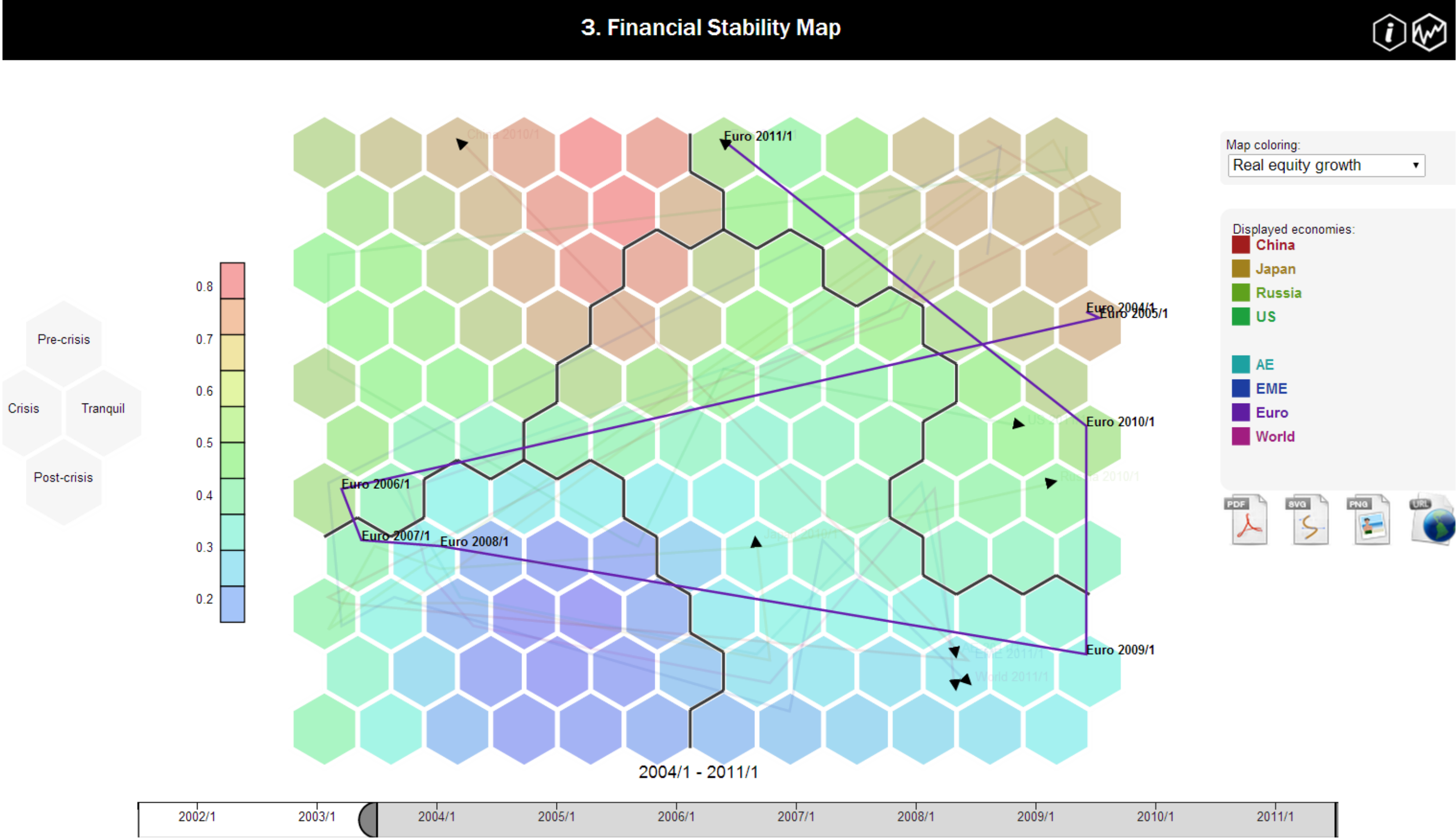}
\par\end{centering}

\textbf{\scriptsize Notes}{\scriptsize : The interactive Financial
Stability Map can be found here: \href{http://vis.risklab.fi/}{http://vis.risklab.fi/}.
In the shown screenshot, the states of financial stability are shown
by location on the map for economies highlighted in the right panel,
and hovering over the label of the euro area dims the time series
for all other economies. The time brush below the figure is used to
visualize trajectory for the chosen economies. The left-hand side
class legend and scale refer to the financial stability states (or
clusters) and the distribution of values for individual indicators,
respectively. These indicators can be used for coloring through the
drop-down menu on the right side.}{\scriptsize \par}

\centering{}\caption{\label{Fig:FSM} An interactive Financial Stability Map.}
\end{figure}

Building upon the above presented FSM application that uses analytical
techniques for visualization, this part brings interactivity into
the picture. An interactive implementation of the FSM can be found
in Figure \ref{Fig:FSM}. The implementation uses the same map as
an \emph{overview}, but goes beyond Figure \ref{Fig:SOM applications}
by enabling various forms of interaction. \emph{Zooming and filtering}
refers to choosing the labels and trajectories to be plotted, including
both the cross-sectional (adjusted through right-side panel) and the
time dimensions (adjusted through time brush and left/right arrows),
rather than showing all labels at once. More over, hovering labels
highlights chosen economies and their trajectories. Further \emph{details
on demand} is provided by the possibility to use the underlying dimensions
or layers of the map for coloring. Through the drop-down menu (and
up/down arrows), the graph shows the distribution of an indicator
or the probability of being a member of one of the financial stability
states with heatmap color coding.

This illustrates the state of individual countries, but how can we
understand the evolution of the entire cross section?

\subsubsection*{Financial Stability Map over Time}

\begin{figure}
\noindent \begin{centering}
(\emph{a})\\
\includegraphics[width=1\columnwidth]{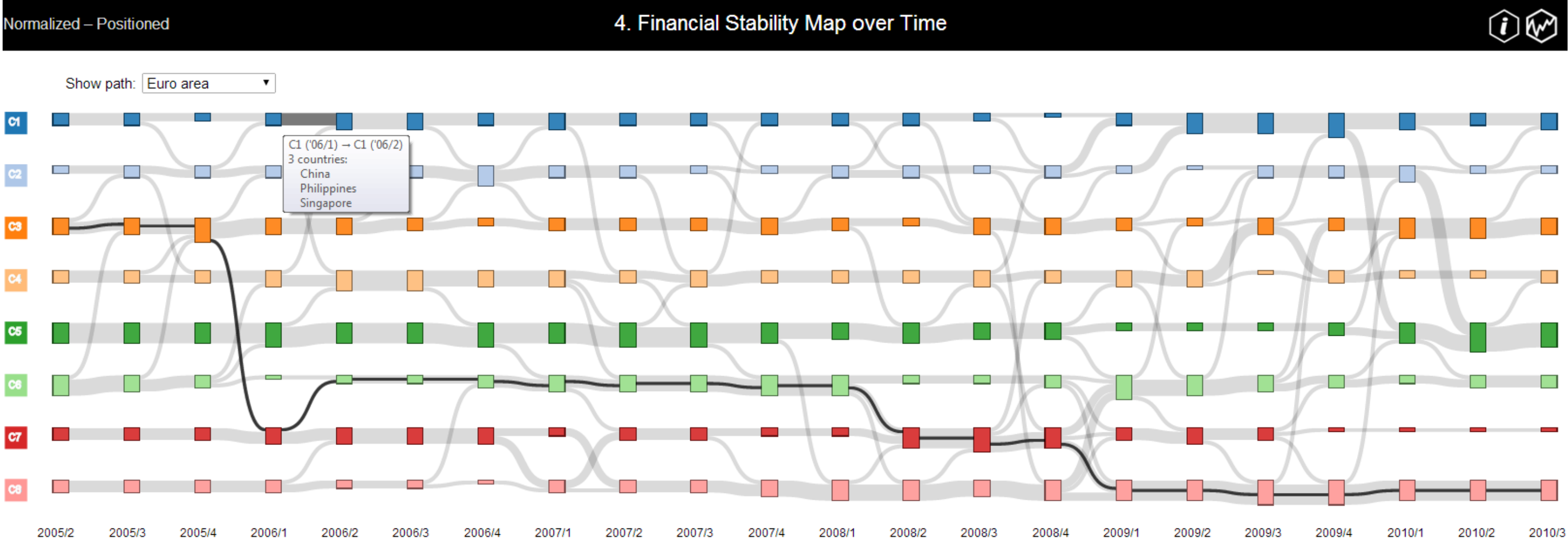}\\
(\emph{b})\emph{}\\
\includegraphics[width=1\columnwidth]{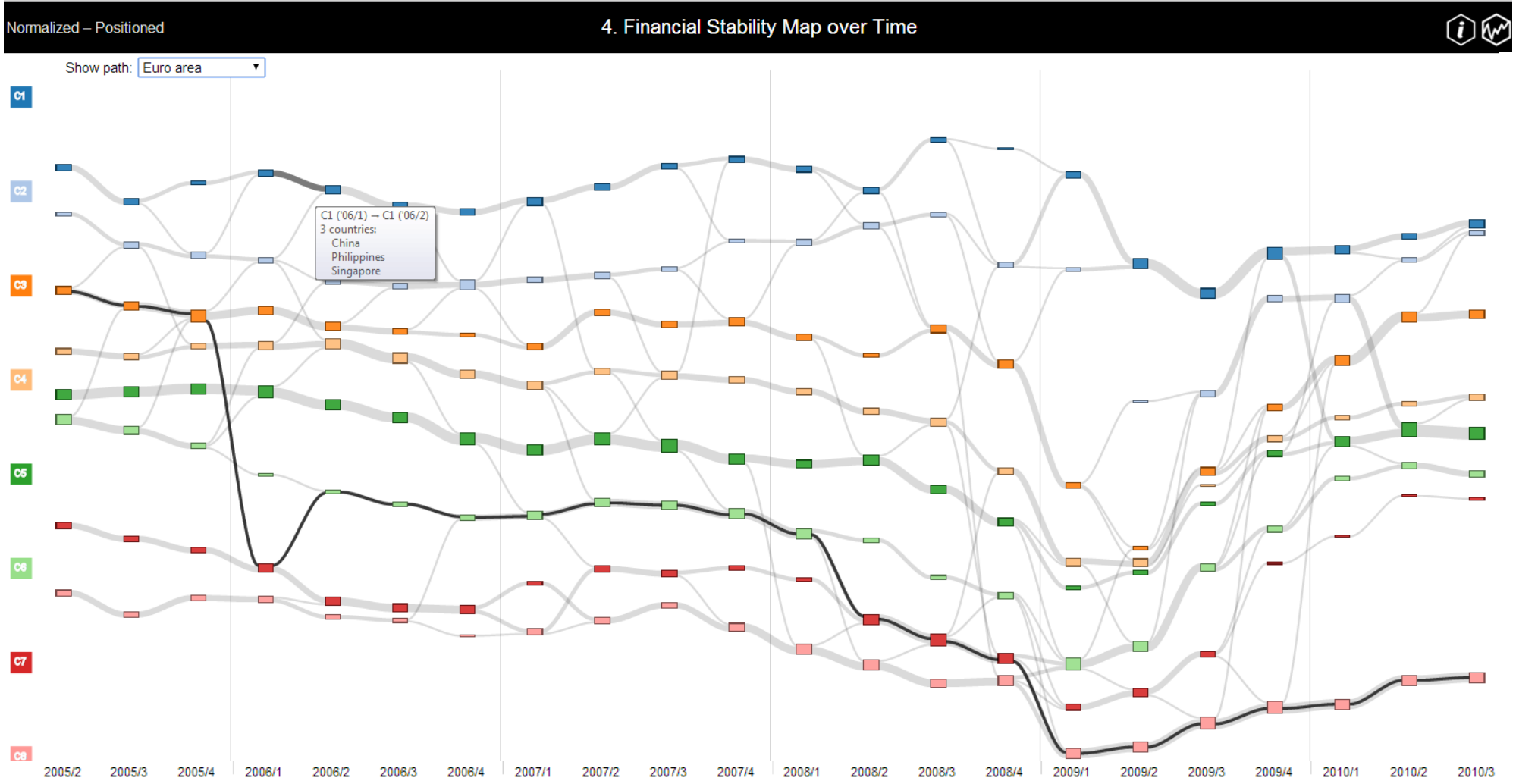}
\par\end{centering}

\textbf{\scriptsize Notes}{\scriptsize : The interactive Financial
Stability Map over Time can be found here: \href{http://vis.risklab.fi/}{http://vis.risklab.fi/}.
In chart (}\emph{\scriptsize a}{\scriptsize ), we illustrate a SOTM
in a standard grid representation and in (}\emph{\scriptsize b}{\scriptsize )
we distort positions to represent changes in cluster structures. Further,
both charts also show a list of economies staying in cluster 1 between
2006Q1 and 2006Q2, as well as the path for the euro area.}{\scriptsize \par}

\centering{}\caption{\label{Fig:FSM-t} An interactive Financial Stability Map over Time.}
\end{figure}

The FSM-t, as above shown, provides means for exploring the evolution
of the cross section, yet it would undeniably benefit from similar
interaction. The visualization design presented herein is based upon
an implementation of alluvial diagrams for representing the SOTM (see
\citealp{RonnqvistSarlin2014a}). As an \emph{overview}, the alluvial
SOTM goes beyond the previous representation by encoding cluster size
by node height, depicting transitions between clusters from a cross
section to the next through links and using a planar visual variable
(\emph{y}-axis) to encode structural changes in data. Figure \ref{Fig:FSM-t}
shows this combined view, where \emph{(a) }shows a standard grid representation
and \emph{(b) }distorts positioning along the vertical dimension.

The implementation supports interaction through various means. Beyond
possibilities to zoom in on an important or dense part and panning
for moving to areas of interest, \emph{zooming and filtering} is supported
through the possibility to drag nodes to better understand linkages
between overlapping or closely neighboring nodes. Also, selecting
individual economies provides means for a more focused view of individual
transitions paths. \emph{Further details on demand} is provided when
hovering over transition links, which provides a list of all switching
economies. One could also see moving from (\emph{a})\emph{ }and (\emph{b})\emph{
}as a an approach to move from a baseline representation to further
details on structural changes.

Now we have provided means for exploring and interacting with the
three more standard dimensions of the data cube, but how could we
interact with graphics of network data?

\subsubsection*{Bank Interrelation Map}

Interaction with network models is an inherently important task. Despite
the popularity of stand-alone force-directed layout algorithms, coupling
them with interaction possibilities, while oftentimes missing, is
essential for not only better data exploration, but also possible
remedies for improving sub-optimal solutions. Moreover, given the
rarity of reaching a global optima, pulling nodes from equilibria,
only to let them migrate back to a new sub-optimal position, increases
our understanding of network properties.

\begin{figure}
\begin{centering}
\includegraphics[width=1\columnwidth]{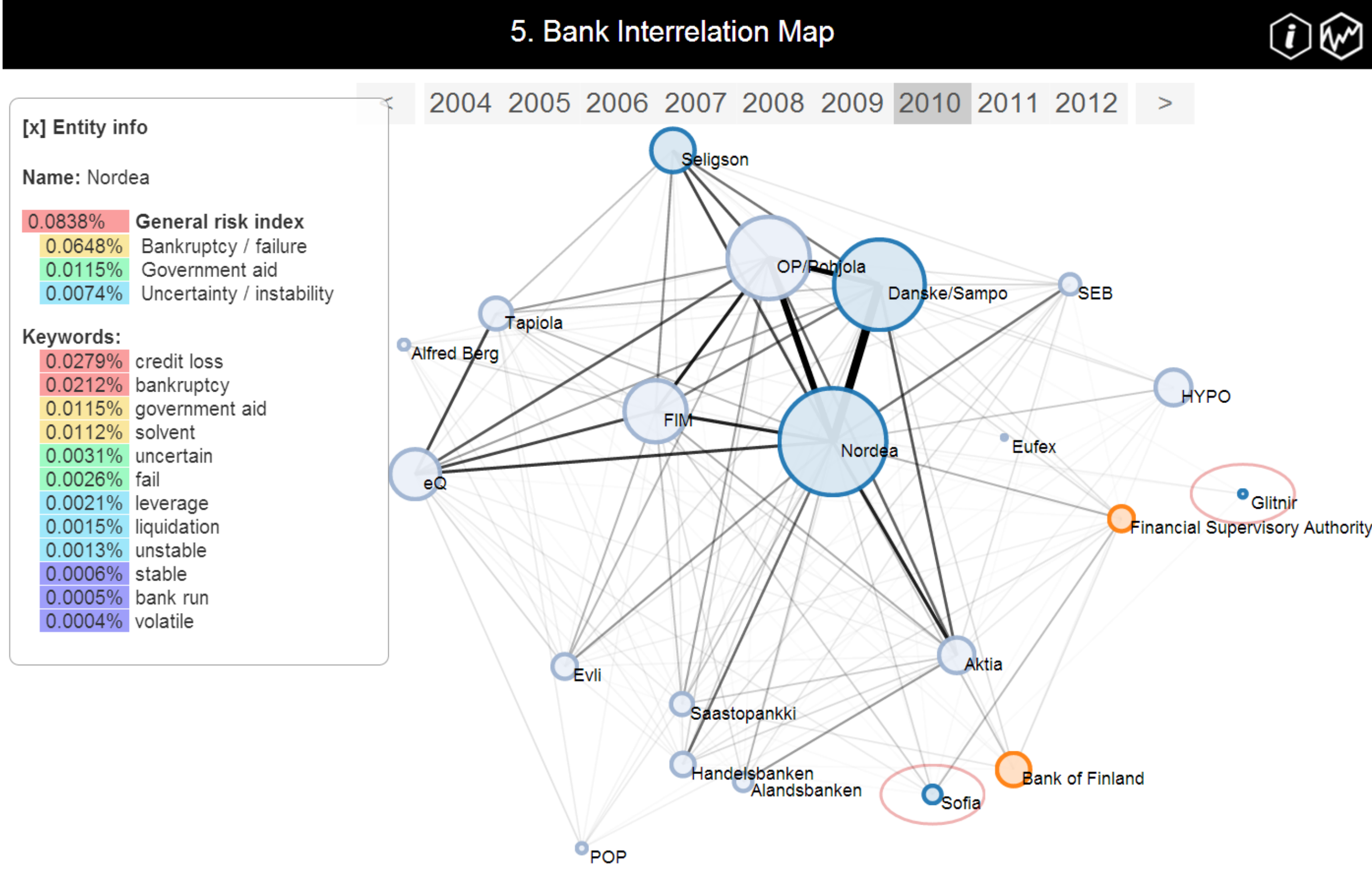}
\par\end{centering}

\textbf{\scriptsize Notes}{\scriptsize : The interactive Bank Interrelation
Map can be found here: \href{http://vis.risklab.fi/}{http://vis.risklab.fi/}.
In the shown screenshot, a network of bank co-occurrences is drawn
with the Fruchterman-Reingold algorithm. Location on the map represents
centrality and overall dependency structure. Highlighting nodes provides
on the left-hand side statistics measuring the share of discussion
relating to risks and distress. The time brush above the figure is
used to choose the time span that the network refers to, as well as
to support exploration of temporal evolution. The red circles indicate
distressed banks during the current crisis (Glitnir, 2008; Sofia;
2010).}{\scriptsize \par}

\centering{}\caption{\label{Fig:BIM} An interactive Bank Interrelation Map.}
\end{figure}

The \emph{overview} of the bank interrelation map is shown in Figure
\ref{Fig:BIM}. \emph{Zooming and filtering} is supported by possibilities
to zoom in on an important or dense part and panning enables moving
to areas of interest, whereas dragging nodes enables both to adjust
the current optimum of the layout algorithm and to alter the orientation
of the graph and to better understand linkages to overlapping or closely
neighboring nodes. Moreover, a time brush enables also varying the
time span to range from all years to one year and exploring the evolution
of the network over time. Further, the up-down arrow keys enable filtering
out the linkages and re-running the algorithm with a random initialization.
The variation caused by different initial values highlights the importance
of interaction, as two suboptimal solutions may significantly differ.
While the richness of textual data would enable a wide range of various
further queries, in this application we showcase the feature of \emph{details
on demand} by coupling highlighting of nodes with a panel on the left
side showing the share of discussion relating to risks and distress.

The use of layout algorithms, while providing means to low-dimensional
representations of complex data, is no panacea for visualization.
In particular, they lack general-purpose solutions to representing
densely connected networks and populated areas. To move beyond the
exploration features that are enabled by interaction, layout algorithms
can be coupled with other approaches. Entirely different methods may
in some cases provide better features for simple representation. For
instance, Minimum Spanning Trees provide a simplified skeleton of
large networks. Other approaches facilitating the representation of
cluttered networks include node and dimension grouping and filtering
(see, e.g., \citealp{MaMuelder2013}), circular and chord diagrams
with hierarchical edge bundling (see, e.g., \citealp{Holten2006})
and edge lenses and magnifiers (see, e.g., \citet{Wongetal2003}).
Likewise, given that the data can be split into smaller, meaningful
groups, the Hive Plot \citep{Krzywinskietal2012} provides ample means
for a pure representation of relationships between entities, as well
as in communicating system-wide connectedness.

\subsection{VisRisk: A visualization platform for macroprudential analysis}

So far, this section has presented applications of analytical and
interactive visualizations. Now, we move one step further by putting
forward VisRisk as a general platform for visualizing the macroprudential
data cube. Consisting of analytical and interactive visualizations,
VisRisk comes with three modules: \emph{plots}, \emph{maps} and \emph{networks}.%
\footnote{The VisRisk platform for interactive and analytical applications can
be found here:

\href{http://vis.risklab.fi/}{http://vis.risklab.fi/}%
} \emph{Plots} focus on interactive interfaces for representing large
amounts of data, but do not make use of analytical techniques for
reducing complexity. While \emph{maps} provide analytical means for
representing the three standard dimensions of a data cube in simple
formats, \emph{networks} aim at visualization of the fourth data cube
dimension of interlinkages. In this section, we have illustrated applications
of five web-based interactive visualizations to systemic risk indicators
and models, of which three make use of analytical visualizations.
First, we made use of analytical techniques for data and dimension
reduction to explore high-dimensional systemic risk indicators and
time-varying networks of linkages. Second, we added interactivity
to not only dashboards of standard risk indicators and early-warning
models, but also to the analytical applications. This spanned the
spectrum of the three modules: \emph{plots}, \emph{maps} and \emph{networks}.
Table \ref{tab:The-platform-modules} summarizes the relations of
the applications to the modules and analytical and interactive features.

\begin{table}
\caption{The applications in relation to the platform and analytical and interactive
visualizations\label{tab:The-platform-modules}}

\begingroup\tabcolsep=3pt\def\arraystretch{1.7}

\noindent \begin{centering}
\begin{tabular}{c>{\centering}p{7cm}>{\centering}p{5cm}}
 & \textbf{Analytical visualization} & \textbf{Interactive visualization}\tabularnewline
\hline 
\hline 
\emph{plots} &  & 1. Risk dashboard

2. Early-warning model\tabularnewline
\emph{maps} & 3. FSM

4. FSM-t & 3. FSM

4. FSM-t\tabularnewline
\emph{networks} & 5. BIM & 5. BIM\tabularnewline
\end{tabular}
\par\end{centering}

\endgroup
\end{table}

From the viewpoint of the data cube in Figure \ref{Fig:financialdatacube},
VisRisk's three modules provide visual means to explore all four dimensions.
The tasks of the modules can be described as follows:
\begin{enumerate}
\item \emph{plots}: To make use of interactive interfaces to standard two-dimensional
graphs for representing large amounts of data from the macroprudential
data cube without the use of any analytical approaches for reducing
complexity.
\item \emph{maps}: To couple analytical data and dimension reduction techniques
with interactive visualizations for reducing complexity of the three
standard dimensions of the macroprudential data cube, in order to
represent large-volume, high-dimensional and/or high-frequency data
in simpler formats.
\item \emph{networks}: To couple analytical data and dimension reduction
techniques with interactive visualizations for reducing complexity
of the fourth, network dimension of the macroprudential data cube,
in order to represent large-volume, multi-layer and/or time-varying
networks in simpler formats.
\end{enumerate}
Beyond the applications herein, the ultimate aim of this paper is
to provide VisRisk as a platform or basis for the use of visualization
techniques, especially those including analytical and interactive
features, in macroprudential oversight in general and risk communication
in particular. Hence, the platform enables and is open to the visualization
of any data from the macroprudential data cube. This aims at supporting
the understanding and value of analytical and interactive visualizations,
in addition to which the consolidation of systemic risk indicators
and models can be seen as a support for assessing and comparing systemic
risk indicators and models. Possibilities to graphically explore and
compare a wide variety of risk measures strives to broadly support
macroprudential analysis and the development of new measures.

\section{Conclusions}

Macroprudential oversight concerns surveillance and supervision of
the financial system as a whole. This paper has brought the topic
of visualization to the discussion of risk communication in macroprudential
oversight. Communicating timely information related to systemic risks
broadly and effectively is a key mandate of macroprudential supervisory
bodies. Likewise, while the mandate of multiple macroprudential supervisors
imposes a need to analyze a large number of entities and their constituents
as a whole, the soar in availability and precision of data further
stresses the importance of simple representations of complex data.

To address tasks of big data and communication in macroprudential
oversight, it becomes evident that visual interfaces hold promise
as a means for supporting policymakers. Indeed, visual interfaces
are already today essential in supporting everyday decisions, ranging
from visuals in human-computer interaction to the standard set of
two-dimensional graphs of statistics used in external communication.
This paper takes a step further by matching the tasks of a macroprudential
supervisor with visualization techniques available today, to achieve
maximum usefulness of available data. To support the use of big data
and analytical tools for timely and accurate measurement of systemic
risk, one essential ingredient to dealing with complex data and modeling
problems is to improve end users' understanding of them. A particular
benefit relates, as noted by Jean-Claude Trichet, to the support of
\textit{\emph{disciplined and structured judgmental analysis}} based
upon policymakers' experience and domain intelligence. Further, the
mandates of macroprudential supervisors most often stress, or are
even limited to, issuing risk warnings and policy recommendations,
as well as overall communication. This highlights the importance of
communicating broadly and effectively timely information about systemic
risks.

This paper has discussed the fields of information visualization and
visual analytics, as well as techniques provided within them, as potential
means for risk communication. Particularly, a common thread throughout
the paper has been to draw upon the visualization literature, in order
to better support the tasks of macroprudential oversight. We have
concluded that two essential features for supporting the analysis
of big data and communication of risks are analytical visualizations
and interactive interfaces. 

For visualizing the macroprudential data cube through analytical and
interactive visualization, this paper has provided the VisRisk platform
with three modules: \emph{plots}, \emph{maps} and \emph{networks}.
We have illustrated the platform and its modules with five web-based
interactive visualizations of systemic risk indicators and models,
of which three make use of analytical visualizations. As VisRisk enables
and is open to the visualization of any data from a macroprudential
data cube, the work in this paper aims at providing a basis with which
systemic risk indicators and models can be widely communicated. The
illustrative applications highlight the usefulness of coupling analytical
and interactive visualizations with systemic risk indicators and models,
which VisRisk brings into web-based implementations to supplement
the static view of today's notion of a document. The aim is to change
this view, by advocating the use of interactive data-driven documents
and analytical visualization techniques -- both with an ultimate aim
of improved means for disseminating information.

\newpage{}

\section*{\textmd{\small \renewcommand\refname{References}\bibliographystyle{elsarticle-harv}
\phantomsection\addcontentsline{toc}{section}{\refname}\bibliography{References/references}

\begin{thebibliography}{81}
\expandafter\ifx\csname natexlab\endcsname\relax\def\natexlab#1{#1}\fi
\expandafter\ifx\csname url\endcsname\relax
  \def\url#1{\texttt{#1}}\fi
\expandafter\ifx\csname urlprefix\endcsname\relax\def\urlprefix{URL }\fi

\bibitem[{Allen et~al.(2004)Allen, Francke, and Swinburne}]{Allenetal2004}
Allen, F., Francke, L., Swinburne, M., 2004. Assessment of the riksbank's work
  on financial stability issues, {Sveriges Riksbank Economic Review 3/2004}.

\bibitem[{Arciniegas~Rueda and Arciniegas(2009)}]{ArciniegasRuedaA09}
Arciniegas~Rueda, I., Arciniegas, F., 2009. {SOM}-based data analysis of
  speculative attacks' real effects. Intelligent Data Analysis 13(2), 261--300.

\bibitem[{Armstrong(1985)}]{Armstrong1985}
Armstrong, J., 1985. Long-Range Forecasting, from Crystal Ball to Computer. New
  York: John Wile \& Sons Inc.

\bibitem[{Baddeley and Logie(1999)}]{BaddeleyLogie1999}
Baddeley, A., Logie, R., 1999. Working memory: The multiple-component model.
  In: Miyake, A., Shah, P. (Eds.), Models of Working Memory. Cambridge
  University Press, New York, pp. 28--61.

\bibitem[{Bech and Atalay(2010)}]{BechAtalaya2010}
Bech, M., Atalay, E., 2010. The topology of the federal funds market. Physica
  A: Statistical Mechanics and its Applications 389(22), 5223--5246.

\bibitem[{Bertin(1983)}]{Bertin1983}
Bertin, J., 1983. Semiology of Graphics. The University of Wisconsin Press, WI.

\bibitem[{BIS(1986)}]{BIS1986}
BIS, 1986. Recent innovations in international banking, {R}eport prepared by a
  Study Group established by the central banks of the Group of Ten countries,
  Basel, April.

\bibitem[{Bisias et~al.(2012)Bisias, Flood, Lo, and Valavanis}]{Bisiasetal2012}
Bisias, D., Flood, M., Lo, A., Valavanis, S., 2012. A survey of systemic risk
  analytics. Annual Review of Financial Economics 4, 255--296.

\bibitem[{Borio(2009)}]{Borio2009}
Borio, 2009. Implementing the macroprudential approach to financial regulation
  and supervision. In: Banque de France Financial Stability Review No. 13
  (December 2009).

\bibitem[{Borio(2011)}]{Borio2011}
Borio, C., 2011. Implementing a macroprudential framework: Blending boldness
  and realism. Capitalism and Society 6(1).

\bibitem[{Born et~al.(2013)Born, Ehrmann, and Fratzscher}]{Bornetal2013}
Born, B., Ehrmann, M., Fratzscher, M., 2013. Central bank communication on
  financial stability. The Economic Journal forthcoming.

\bibitem[{Boss et~al.(2006)Boss, Krenn, Puhr, and Summer}]{Bossetal2006}
Boss, M., Krenn, G., Puhr, C., Summer, M., 2006. Systemic risk monitor: A model
  for systemic risk analysis and stress testing of banking systems.
  Oesterreichische Nationalbank Financial Stability Report 11(June), 83--95.

\bibitem[{Bundesbank(2013)}]{Buba2013}
Bundesbank, 2013. Macroprudential oversight in germany: Framework, institutions
  and tools, {Deutsche Bundesbank, Monthly Report, April 2013, pp. 39--54}.

\bibitem[{Card et~al.(1999)Card, Mackinlay, and Schneidermann}]{Cardetal1999}
Card, S., Mackinlay, J., Schneidermann, B., 1999. Readings in information
  visualization, Using Vision to Think. Academic Press Inc., San Diego, CA.

\bibitem[{Card et~al.(1991)Card, Robertson, and Mackinlay}]{Cardetal1991}
Card, S., Robertson, G., Mackinlay, J., 1991. The information visualizer, an
  information workspace. In: Proceedings of CHI '91, ACM Conference on Human
  Factors in Computing Systems. New Orleans, pp. 181--188.

\bibitem[{Carletti(2008)}]{Carletti2008}
Carletti, E., 2008. Competition and regulation in banking. In: Boot, A.,
  Thakor, A. (Eds.), Handbook in Financial Intermediation. Elsevier, North
  Holland, pp. 449--482.

\bibitem[{Cihák(2006)}]{Cihak2006}
Cihák, M., 2006. How do central banks write on financial stability?, {IMF
  Working Paper} No. 06/163.

\bibitem[{Dattels et~al.(2010)Dattels, McCaughrin, Miyajima, and
  Puig}]{Dattels2010}
Dattels, P., McCaughrin, R., Miyajima, K., Puig, J., 2010. Can you map global
  financial stability?, {I}MF Working Paper No. 10/145.

\bibitem[{de~Bandt and Hartmann(2002)}]{deBandtHartmann2002}
de~Bandt, O., Hartmann, P., 2002. Systemic risk in banking: A survey. In:
  Goodhart, C., Illing, G. (Eds.), Financial crisis, contagion and the lender
  of last resort: A book of readings. Oxford University Press, Oxford.

\bibitem[{de~Bandt et~al.(2009)de~Bandt, Hartmann, and Peydro}]{deBandt2009}
de~Bandt, O., Hartmann, P., Peydro, J., 2009. Systemig risk in banking: An
  update. In: Berger, A., M.~P., Wilson, J. (Eds.), Oxfoord Handbook of
  Banking. Oxford University Press, Oxford.

\bibitem[{Dix et~al.(2004)Dix, Finlay, Abowd, and Beale}]{Dixetal2004}
Dix, A., Finlay, J., Abowd, G., Beale, R., 2004. Human-Computer Interaction,
  3rd Edition. Prentice-Hall, London.

\bibitem[{ECB(2009)}]{ECB2009}
ECB, 2009. The concept of systemic risk. In: Financial Stability Review
  (December 2009). European Central Bank, Frankfurt, Germany.

\bibitem[{ECB(2010)}]{ECB2010}
ECB, 2010. Analytical models and tools for the identification and assessment of
  systemic risks. In: Financial Stability Review (June 2010). European Central
  Bank, Frankfurt, Germany.

\bibitem[{Ekholm(2012)}]{Ekholm2012}
Ekholm, K., 2012. Macroprudential policy and clear communication contribute to
  financial stability, {S}peech to the Swedish Banker's Association, 30 March.

\bibitem[{Estrada(2011)}]{Estrada2011}
Estrada, E., 2011. The Structure of Complex Networks: Theory and Applications.
  Oxford University Press, Oxford.

\bibitem[{Fekete et~al.(2008)Fekete, van Wijk, Stasko, and
  North}]{Feketeetal2008}
Fekete, J.-D., van Wijk, J., Stasko, J., North, C., 2008. The value of
  information visualization. In: Information Visualization: Human-Centered
  Issues and Perspectives. Springer, pp. 1--18.

\bibitem[{Flood and Mendelowitz(2013)}]{FloodMendelowitz2009}
Flood, M., Mendelowitz, A., 2013. Monitoring financial stability in a complex
  world. In: Lemieux, V. (Ed.), Financial Analysis and Risk Management Data
  Governance, Analytics and Life Cycle Management. Springer-Verlag, Heidelberg,
  pp. 15--45.

\bibitem[{Fruchterman and Reingold(1991)}]{FruchtermanReingold1991}
Fruchterman, T., Reingold, E., 1991. Graph drawing by force-directed placement.
  Software: Practice and Experience 21(11), 1129--1164.

\bibitem[{Haroz and Whitney(2012)}]{HarozWhitney2012}
Haroz, S., Whitney, D., 2012. How capacity limits of attention influence
  information visualization effectiveness. IEEE Transactions on Visualization
  and Computer Graphics 18(12), 2402--2410.

\bibitem[{Havre et~al.(2000)Havre, Hetzler, and Nowell}]{Havreetal2000}
Havre, S., Hetzler, B., Nowell, L., 2000. Themeriver: Visualizing theme changes
  over time. In: Proceedings of the IEEE Symposium on Information
  Visualization. pp. 115--123.

\bibitem[{Holten(2006)}]{Holten2006}
Holten, D., 2006. Hierarchical edge bundles: Hierarchical edge bundles. IEEE
  Transactions on Visualization and Computer Graphics 12(5), 741--748.

\bibitem[{Hu(2006)}]{Hu2006}
Hu, Y., 2006. Efficient, high-quality force-directed graph drawing. The
  Mathematica Journal 10(1), 37--71.

\bibitem[{Jain(2010)}]{Jain2010}
Jain, A., 2010. Data clustering: 50 years beyond k-means. Pattern Recognition
  Letters 31(8), 651--666.

\bibitem[{Kaser and Lemire(2007)}]{KaserLemire2007}
Kaser, O., Lemire, D., 2007. Tag-cloud drawing: Algorithms for cloud
  visualization. In: Proceedings of the Tagging and Metadata for Social
  Information Organization Workshop. Banff, Alberta, Canada.

\bibitem[{Keim(2001)}]{Keim2001}
Keim, D., 2001. Visual exploration of large data sets. Communications of the
  ACM 44(8), 38--44.

\bibitem[{Keim et~al.(2010)Keim, Kohlhammer, Ellis, and
  Mannsmann}]{Keimetal2010}
Keim, D., Kohlhammer, J., Ellis, G., Mannsmann, F., 2010. Mastering the
  Information Age. Solving Problems with Visual Analytics. Eurographics
  Association.

\bibitem[{Keim and Kriegel(1996)}]{KeimKriegel1996}
Keim, D., Kriegel, H.-P., 1996. Visualization techniques for mining large
  databases: A comparison. IEEE Transactions on Knowledge and Data Engineering
  8(6), 923--938.

\bibitem[{Keim et~al.(2006)Keim, Mansmann, Schneidewind, and
  Ziegler}]{Keimetal2006}
Keim, D., Mansmann, F., Schneidewind, J., Ziegler, H., 2006. Challenges in
  visual data analysis. In: Proceedings of the IEEE International Conference on
  Information Visualization (iV 13). IEEE Computer Society, London, UK, pp.
  9--16.

\bibitem[{Keim et~al.(2009)Keim, Mansmann, and Thomas}]{Keimetal2009}
Keim, D., Mansmann, F., Thomas, J., 2009. Visual analytics: How much
  visualization and how much analytics? SIGKDD Explorations 11(2), 5--8.

\bibitem[{Kindleberger(1978)}]{Kindleberger1996}
Kindleberger, C., 1978. Manias, Panics, and Crashes: A History of Financial
  Crises. John Wiley \& Sons, New York.

\bibitem[{Koffa(1935)}]{Koffa1935}
Koffa, K., 1935. Principles of Gestalt Psychology. Routledge \& Kegan Paul Ltd,
  London.

\bibitem[{Kohonen(1982)}]{Kohonen1982}
Kohonen, T., 1982. Self-organized formation of topologically correct feature
  maps. Biological Cybernetics 43, 59--69.

\bibitem[{Kohonen(2001)}]{Kohonen2001}
Kohonen, T., 2001. Self-Organizing Maps, 3rd Edition. Springer-Verlag, Berlin.

\bibitem[{Korhonen(2013)}]{Korhonen2013}
Korhonen, P., 2013. Do macroprudential tools require micro-data?, {Irving
  Fisher Committee. "Proceedings of the Porto Workshop on "Integrated
  management of micro-databases", 20-22 June 2013," IFC Bulletins, Bank for
  International Settlements, number 37, December}.

\bibitem[{Krzywinski et~al.(2012)Krzywinski, Birol, Jones, and
  Marra}]{Krzywinskietal2012}
Krzywinski, M., Birol, I., Jones, S., Marra, M., 2012. Hive plots -- rational
  approach to visualizing networks. Briefings in Bioinformatics 3(5), 627--644.

\bibitem[{Larkin and Simon(1987)}]{LarkinSimon1987}
Larkin, J., Simon, H., 1987. Why a diagram is (sometimes) worth ten thousand
  words. Cognitive Science 11, 65--99.

\bibitem[{Lee and Verleysen(2007)}]{LeeVerleysen2007}
Lee, J., Verleysen, M., 2007. Nonlinear dimensionality reduction.
  Springer-Verlag, Information Science and Statistics Series., Heidelberg,
  Germany.

\bibitem[{Lin(1997)}]{Lin1997}
Lin, X., 1997. Map displays for information retrieval. Journal of the American
  Society for Information Science 48(1), 40--54.

\bibitem[{Lo~Duca and Peltonen(2013)}]{Duca2012}
Lo~Duca, M., Peltonen, T., 2013. Assessing systemic risks and predicting
  systemic events. Journal of Banking {\&} Finance 37(7), 2183--2195.

\bibitem[{Ma and Muelder(2013)}]{MaMuelder2013}
Ma, K., Muelder, C., 2013. Large-scale graph visualization and analytics.
  Computer 46(7), 39--46.

\bibitem[{Marghescu et~al.(2010)Marghescu, Sarlin, and Liu}]{ISAF:ISAF317}
Marghescu, D., Sarlin, P., Liu, S., 2010. Early-warning analysis for currency
  crises in emerging markets: A revisit with fuzzy clustering. Intelligent
  Systems in Accounting, Finance \& Management 17~(3-4), 143--165.

\bibitem[{McNees(1990)}]{McNees1990}
McNees, S., 1990. The role of judgment in macroeconomic forecasting accuracy.
  International Journal of Forecasting 6(3), 287--299.

\bibitem[{Minsky(1982)}]{Minsky1982}
Minsky, H., 1982. Can "it" Happen Again?: Essays on Instability and Finance.
  M.E. Sharpe, Armonk, N.Y.

\bibitem[{Mitchell(2001)}]{Mitchell2001}
Mitchell, C., 2001. Selling the brand inside. Harvard Business Review 80(1).

\bibitem[{Mohan(2009)}]{Mohan2009}
Mohan, R., 2009. Communications in central banks: A perspective, {Stanford
  Center for International Development, Working Paper No. 408}.

\bibitem[{Oosterloo and de~Haan(2004)}]{Oosterlooa2004}
Oosterloo, S., de~Haan, J., 2004. Central banks and financial stability: a
  survey. Journal of Financial Stability 1(2), 257--273.

\bibitem[{Risch et~al.(2008)Risch, Kao, Poteet, and Wu}]{Rischetal2008}
Risch, J., Kao, A., Poteet, S., Wu, Y.-J., 2008. Text visualization for visual
  text analytics. In: Simoff, S., Böhlen, M., Mazeika, A. (Eds.), Visual Data
  Mining: Theory, Techniques and Tools for Visual Analytics. Springer, pp.
  154--171.

\bibitem[{Rönnqvist and Sarlin(2014{\natexlab{a}})}]{RonnqvistSarlin2014a}
Rönnqvist, S., Sarlin, P., 2014{\natexlab{a}}. Alluvial {SOTM}: Visualizing
  transitions and changes in cluster structure of the {Self-Organizing Time
  Map}. In: Proceedings of the Eurographics Conference on Visualization
  (EuroVis'14). Eurographics Association, Swansea, UK.

\bibitem[{Rönnqvist and Sarlin(2014{\natexlab{b}})}]{RonnqvistSarlin2014}
Rönnqvist, S., Sarlin, P., 2014{\natexlab{b}}. From text to bank interrelation
  maps. In: Proceedings of the IEEE Symposium on Computational Intelligence for
  Financial Engineering \& Economics (CIFEr 14). IEEE Press, London, UK.

\bibitem[{Sarlin(2010)}]{Sarlin2010Currency}
Sarlin, P., 2010. Visual monitoring of financial stability with a
  self-organizing neural network. In: Proceedings of the International
  Conference on Intelligent Systems Design and Applications (ISDA 10). IEEE
  Press, Cairo, Egypt, pp. 248--253.

\bibitem[{Sarlin(2011)}]{Sarlin2011Debt}
Sarlin, P., 2011. Sovereign debt monitor: A visual self-organizing maps
  approach. In: Proceedings of IEEE Symposium on Computational Intelligence for
  Financial Engineering and Economics (CIFEr 11). IEEE Press, Paris, France,
  pp. 1--8.

\bibitem[{Sarlin(2013{\natexlab{a}})}]{SarlinPRL2013}
Sarlin, P., 2013{\natexlab{a}}. Decomposing the global financial crisis: A
  self-organizing time map. Pattern Recognition Letters 34, 1701--1709.

\bibitem[{Sarlin(2013{\natexlab{b}})}]{Sarlin2012}
Sarlin, P., 2013{\natexlab{b}}. Self-organizing time map: An abstraction of
  temporal multivariate patterns. Neurocomputing 99(1), 496--508.

\bibitem[{Sarlin(2014{\natexlab{a}})}]{SarlinIV2013}
Sarlin, P., 2014{\natexlab{a}}. Data and dimension reduction for visual
  financial performance analysis. Information Visualization, forthcoming, doi:
  10.1177/1473871613504102.

\bibitem[{Sarlin(2014{\natexlab{b}})}]{Sarlinthesis2013}
Sarlin, P., 2014{\natexlab{b}}. Mapping Financial Stability. Computational Risk
  Management Series. Springer Verlag.

\bibitem[{Sarlin and Marghescu(2011)}]{ISAF:ISAF321}
Sarlin, P., Marghescu, D., 2011. Visual predictions of currency crises using
  self-organizing maps. Intelligent Systems in Accounting, Finance and
  Management 18(1), 15--38.

\bibitem[{Sarlin and Peltonen(2013)}]{SarlinPeltonen2013}
Sarlin, P., Peltonen, T., 2013. Mapping the state of financial stability.
  Journal of International Financial Markets, Institutions \& Money 26, 46--76.

\bibitem[{Shannon and Weaver(1963)}]{ShannonWeaver1963}
Shannon, C., Weaver, W., 1963. A Mathematical Theory of Communication.
  University of Illinois Press, Champaign.

\bibitem[{Shneiderman(1996)}]{Shneiderman1996}
Shneiderman, B., 1996. The eyes have it: A task by data type taxonomy for
  information visualizations. In: Proceedings of the IEEE Symposium on Visual
  Languages. Boulder, CO, pp. 336--343.

\bibitem[{Thomas and Cook(2005)}]{ThomasCook2005}
Thomas, J., Cook, K., 2005. Illuminating the Path: Research and Development
  Agenda for Visual Analytics. IEEE Press.

\bibitem[{Tikhonova and Ma(2008)}]{TikhonovaMa2008}
Tikhonova, A., Ma, K., April 2008. A scalable parallel force-directed graph
  layout algorithm. In: Proceedings of Parallel Graphics and Visualization
  Symposium.

\bibitem[{Trichet(2009)}]{Trichet2009}
Trichet, J., 2009. Systemic risk, {Clair Distinguised Lecture in Economics and
  Public Policy, Cambridge University, Cambridge, December 2009}.

\bibitem[{Triesman(1985)}]{Triesman1985}
Triesman, A., 1985. Preattentive processing in vision. Computer Vision,
  Graphics, and Image Processing 31(2), 156--177.

\bibitem[{Tufte(1983)}]{Tufte1983}
Tufte, E., 1983. The Visual Display of Quantitative Information. Graphics
  Press, Cheshire, CT.

\bibitem[{Tukey(1977)}]{Tukey1977}
Tukey, J., 1977. Exploratory Data Analysis. Addison-Wesley, Reading, PA.

\bibitem[{von Landesberger et~al.(2014)von Landesberger, Diel, Bremm, and
  Fellner}]{vonLandesberger2014}
von Landesberger, T., Diel, S., Bremm, S., Fellner, D., 2014. Visual analysis
  of contagion in networks. Information Visualization, forthcoming.

\bibitem[{Ware(2004)}]{Ware2004}
Ware, C., 2004. Information Visualization: Perception for Design. Morgan
  Kaufman, San Francisco, CA.

\bibitem[{Ware(2005)}]{Ware2005}
Ware, C., 2005. Visual queries: the foundation of visual thinking. In: Tergan,
  S., Keller, T. (Eds.), Knowledge and information visualization. Springer,
  Berlin, Germany, pp. 27--35.

\bibitem[{Wong et~al.(2003)Wong, Carpendale, and Greenberg}]{Wongetal2003}
Wong, N., Carpendale, S., Greenberg, S., 2003. Edgelens: An interactive method
  for managing edge congestion in graphs. In: Proceedinngs of the IEEE
  Symposium on Information Visualization. IEEE, Los Alamitos, CA, USA, pp.
  51--58.

\bibitem[{Woolford(2001)}]{Woolford2001}
Woolford, I., 2001. Macro-financial stability and macroprudential analysis.
  Reserve Bank of New Zealand Bulletin 64.

\bibitem[{Zhang et~al.(2012)Zhang, Stoffel, Behrisch, Mittelstädt, Schreck,
  Weber, Last, and Keim}]{Zhangetal2012}
Zhang, L., Stoffel, A., Behrisch, M., Mittelstädt, S., Schreck, T.~Pompl, R.,
  Weber, S., Last, H., Keim, D., 2012. Visual analytics for the big data era --
  a comparative review of state-of-the-art commercial systems. In: Proceedings
  of the IEEE Conference on Visual Analytics Science and Technology (VAST).
  Seattle, WA, pp. 173--182.

\end{thebibliography}
}}

\newpage{}

\section*{Appendix A: Information visualization and visual analytics}

This appendix provides supplementary information regarding topics
in information visualization and visual analytics.

\subsection*{Appendix A.1: How visuals amplify cognition}

\citet{Cardetal1999} present visuals as a means to amplify cognition
and list five ways how well-perceived visuals could amplify cognition.
They can be exemplified as follows.

Examples of the \emph{first} way to amplify cognition, the increase
in available resources, are parallel perceptual or visual processing
and offl{}oading work from the cognitive system to the perceptual
system \citep{LarkinSimon1987}. \emph{Second}, visuals facilitate
the search procedure by the provision of a large amount of data in
a small space (i.e., high data density) \citep{Tufte1983} and by
grouping information used together in general and information about
one object in particular \citep{LarkinSimon1987}. \emph{Third}, abstraction
and aggregation aid in the detection of patterns and operations for
perceptual inference \citep{Cardetal1991}. \emph{Fourth}, perceptual
monitoring is enhanced, for instance, through the use of pop-out effects
created by appearance or motion \citep{Cardetal1999}. Likewise, \citet{Cardetal1999}
exemplify the \emph{fifth} way to amplify cognition, the use of a
manipulable medium, by allowing the user to explore a wide range of
parameter values to interactively explore properties of data.

\subsection*{Appendix A.2: The visual system and correctly perceived graphics}

Visual representations, while providing means to amplify cognition,
also constitute a large set of issues that may hinder, disturb or
otherwise negatively affect how visualizations are read. A key starting
point is to take into account the deficiencies and limitations of
human perception. Preattentive processing, for instance, becomes a
deficiency if visuals are not designed properly. Patterns a user is
supposed to identify quickly -- or give visual but not conscious attention
to -- should hence be made distinct from the rest by using features
that can be preattentively processed. Likewise, visual attention functions
as a filter in that only one pattern is brought into working memory
\citep{BaddeleyLogie1999}. Hence, if provided with multiple patterns,
we only see what we need or desire to see by tuning out other patterns.
It is also important to note differences in visual features, as others
are perceived more accurately, such as color vs. position, where perception
of the latter dominates over the former. \citet{Ware2005} also mentions
the fact that humans process simple visual patterns serially at a
rate of one every 40--50 msec. and a fixation lasts for about 100--300
msec., meaning that our visual system processes 2--6 objects within
each fixation, before we move our eyes to visually attend to some
other region. In addition, one important factor to account for is
how perception of visuals is affected by properties of the human eye,
such as acuities, contrast sensitivity, color vision, perception of
shape or motion with colors, etc. Another aspect of crucial importance
is obviously to pay regard to human perceptions of shapes in visuals,
such as distances, sizes and forms. Cognitive deficiencies should
also be accounted for when designing visuals, such as the limited
amount of working memory. For instance, \citet{HarozWhitney2012}
show that the effectiveness of information visualization is severely
affected by the capacity limits of attention, not the least for detecting
unexpected information. Hence, an understanding of the functioning
of the human visual system aids in producing effective displays of
information, where data are presented such that the patterns are likely
to be correctly perceived.

\subsection*{Appendix A.3: A framework of the planar and retinal variables}

\citet{Bertin1983} describes the \emph{plane}, and its two dimensions
$(x,y)$, as the richest variables. They fulfill the criteria for
all levels of organization by being selective, associative, ordered
and quantitative. The retinal variables, on the other hand, are always
positioned on the plane, and can make use of three types of implantation:
a point, line, or area. First, \emph{size} is ordered, selective but
not associative, and the only quantitative retinal variable. Second,
\emph{value} is the ratio of black to white on a surface, according
to the perceived ratio of the observer, and is also sometimes called
brightness. The usage of value in this case is close to that in the
HSV (hue, saturation and value) color space, which is a cylindrical-coordinate
representation of points in an RGB (red, green and blue) color space.
It is an ordered and selective retinal variable. Third, \emph{texture}
represents the scale of the constituent parts of a pattern, where
variation in texture may occur through photographic reductions of
a pattern of marks. That is, it may range from null texture with numerous
but tiny elements that are not identifiable to large textures with
only few marks. Texture as a retinal variable can be ordered and is
both selective and associative. Fourth, variation may occur in \emph{color}.
The variation of two marks with the same value or brightness is thus
more related to changes in hue of HSV. Color as a retinal variable
is selective and associative, but not ordered. Fifth, the \emph{orientation}
variable enables variation in the angle between marks. In theory,
this opens up an infinite set of alternatives of the available 360
degrees, whereas \citet{Bertin1983} suggests the use of four steps
of orientation. The orientation variable is associative and selective
only in the cases of points and lines, but has no direct interpretation
of order. Finally, the sixth variable of \emph{shape}, while being
a retinal variable on its own, also partly incorporates aspects of
size and orientation. It is associative, but neither selective nor
ordered.

Tufte's \citeyearpar{Tufte1983} two principles on graphical excellence
and integrity covered in the main text, while being his main guidelines
on graphic design, cover only a small fraction of his work. Beyond
these two principles, but still relating to them, he highlights \emph{data-ink
maximization}, by advocating a focus on the data, and nothing else.
Hence, a good graphical representation focuses on data-ink maximization
with minimum non-data-ink. The data-ink ratio is calculated by 1 minus
the proportion of the graph that can be erased without loss of data
information. Tufte puts forward the following five guidelines related
to data ink: \emph{i}) above all else, show data; \emph{ii}) maximize
the data-ink ratio; \emph{iii}) erase non-data-ink; \emph{iv}) erase
redundant data-ink; and \emph{v}) revise and edit. Moreover, \citet{Tufte1983}
highlights \emph{data density maximization}, which relates to the
share of the area of the graphic dedicated to showing the data. For
too low densities, graphics should either be reduced in size (shrink
principle) or replaced with a table. In terms of concrete design,
he proposes the small multiples, a design for showing varying data
onto a series of the same small graph repeated in one visual.

\subsection*{Appendix A.4: Visualization techniques as per data and output}

\citet{Zhangetal2012} categorize visualization techniques into four
groups from the viewpoint of the underlying data. First, \emph{numerical
data} can be visualized by a vast number of approaches, such as standard
visualization techniques like bar and pie charts and scatter plots.
These focus most often on the visualization of low-dimensional numerical
data. On the other hand, visualization techniques like parallel coordinates,
heatmaps and scatter plot matrices provide means to display data with
higher dimensionality. Second, visualization of \emph{textual data}
is a relatively new but rapidly growing fi{}eld. Recent well-known
techniques include theme river \citep{Havreetal2000} and word cloud
\citep{KaserLemire2007}, for instance. Likewise, the availability
of the third type of data, \emph{geo-tagged data}, has caused a soar
in the demand for geo-spatial visualizations. Geo-related univariate
or multivariate information is oftentimes projected into conventional
two-dimensional and three-dimensional spaces. Fourth, graph visualizations
provide means for displaying patterns in \emph{network data} with
relationships (i.e., edges) between entities (i.e., nodes). They most
often consist of a technique for positioning, such as force-directed
drawing algorithms, as well as coloring or thickness of edges to display
the size of a relationship. Graph or network visualizations have been
increasingly applied in a wide range of emerging topics related to
social and biological networks, not to mention financial networks.

From the viewpoint of visualization output, \citet{KeimKriegel1996}
provide a five-category grouping of techniques. First, \emph{geometric
techniques} provide means for visualization of geometric transformations
and projections of data. Examples of the methods are scatterplot-matrices,
parallel-coordinate plots and projection methods. Second, \emph{icon-based
techniques}, as already the name states, visualize data as features
of icons. The methods include, for instance, Chernoff-faces and stick
figures, of which the former visualize multidimensional data using
the properties of a face icon and the latter use stick figures. Third,
\emph{pixel-oriented techniques} map each attribute value to a colored
pixel and present attribute values belonging to each attribute in
separate subwindows. For instance, query-independent techniques arrange
data from top-down in a column-by-column fashion or left to right
in a line-by-line fashion, while query-dependent techniques visualize
data in the context of a specific user query. Four, \emph{hierarchical
techniques} provide means to illustrate hierarchical structures in
data. Most often, hierarchical methods focus on dividing an $n$-dimensional
attribute space by \textquoteleft{}stacking\textquoteright{} two-dimensional
subspaces into each other. Finally, the fifth category, \emph{graph-based
techniques}, focus on the visualization of large graphs, or networks,
to illustrate the properties of the network, as was above discussed.
In addition, Keim and Kriegel also illustrate the existence of a wide
range of hybrids that make use of multiple categories.

\newpage{}

\section*{Appendix B: Analytical techniques}

This appendix provides supplementary technical details regarding the
analytical methods applied in the paper.

\subsection*{Appendix B.1: Self-Organizing Map}

The \ac{SOM} \citep{Kohonen1982,Kohonen2001} is a method that performs
a simultaneous data and dimension reduction. It differs from non-linear
projection techniques like \ac{MDS} by attempting to preserve the
neighborhood relations in a data space $\Omega$ on a $k$-dimensional
array of units (represented by reference vectors $m_{i}$) instead
of attempting to preserve absolute distances in a continuous space.
On the other hand, it differs from standard \ac{VQ} by also attempting
neighborhood preservation of $m_{i}$. The \ac{VQ} capability of
the \ac{SOM} performs this data reduction into mean profiles (i.e.,
units $m_{i}$). It models from the continuous space $\Omega$, with
a probability density function $p(x)$, to the grid of units, whose
location depend on the neighborhood structure of the data $\Omega$.

We employ the batch training algorithm, and thus process data simultaneously
instead of in sequences. This reduces computational cost and enables
reproducible results. Following an initialization based upon two principal
components, the batch training algorithm operates a specified number
of iterations $t$ (where $t=1,2,\lyxmathsym{\ldots},T$) in two steps.
In the first step, each input data vector $x_{j}$ is assigned to
the \acp{BMU} $m_{b}$:

\begin{equation}
d_{x}(j,b)=\min_{i}d_{x}(j,i),
\end{equation}

where $d_{x}(j,b)$ is the input space distance between data $x_{j}$
and reference vector $m_{b}$ (i.e., \ac{BMU}) and $d_{x}(j,i)$
is the input space distance between data $x_{j}$ and each reference
vector $m_{i}$. Hence, data are projected to an equidimensional reference
vector $m_{b}$, not a two-dimensional vector as in \ac{MDS}. In
the second step, each reference vector $m_{i}$ (where $i=1,2,\lyxmathsym{\ldots},M$)
is adjusted using the batch update formula:

\begin{equation}
m_{i}(t+1)=\frac{\sum_{j=1}^{N}h_{ib(j)}(t)x_{j}}{\sum_{j=1}^{N}h_{ib(j)}(t)}
\end{equation}

where index $j$ indicates the input data vectors that belong to unit
$b$, $N$ is the number of the data vectors, and $h_{ib(j)}$ is
some specified neighborhood function. In comparison to the update
formula of the $k$-means algorithm, the batch update of the \ac{SOM}
can be seen as a spatially $\left(h_{ib(j)}\right)$ constrained version.
The neighborhood function $h_{ib(j)}\in(0,1]$ is defined as the following
Gaussian function: 

\begin{equation}
h_{ib(j)}=\exp\left(-\frac{d_{r}(b,i)^{2}}{2\sigma^{2}(t)}\right)
\end{equation}

where $d_{r}(b,i)$ is the distance between the coordinates $r_{b}$
and $r_{i}$ of the reference vectors $m_{b}$ and $m_{i}$ on the
two-dimensional grid. Moreover, the radius of the neighborhood $\sigma(t)$
is a monotonically decreasing function of time $t$. The radius of
the neighborhood begins as half the diagonal of the grid size ($(X^{2}+Y^{2})/2$),
and decreases towards a user-specified radius $\sigma$.

\subsection*{Appendix B.2: Self-Organizing Time Map}

The \ac{SOTM} \citep{Sarlin2012} uses the clustering and projection
capabilities of the standard \ac{SOM} for visualization and abstraction
of temporal structural changes in data. Here, $t$ (where $t=1,2,\text{\ldots},T$)
is a time-coordinate in data, not in training iterations as is common
for the standard SOM. To observe the cross-sectional structures of
the dataset for each time unit $t$, the \ac{SOTM} performs a mapping
from the input data space $\Omega(t)$, with a probability density
function $p(x,t)$, onto a one-dimensional array $A(t)$ of output
units $m_{i}(t)$ (where $i=1,2,\text{\ldots},M$). Preservation of
orientation and gradual adjustment to temporal changes is accomplished
by initializing $A(t_{1})$ with the first principal component of
\ac{PCA} and initializing A($t_{2,3,...,T}$) with the reference
vectors of $A(t-1)$. Hence, the model uses short-term memory to retain
information about past patterns and preserve orientation. Adjustment
to temporal changes is achieved by performing a batch update per time
$t$. For $A(t_{1,2,\text{\ldots},T})$, each data point $x_{j}(t)\in\Omega(t)$
(where $j=1,2,\lyxmathsym{\ldots},N(t)$) is compared to reference
vectors $m_{i}(t)\in A(t)$ and assigned to its \ac{BMU} $m_{b}(t)$: 

\begin{equation}
\left\Vert x_{j}(t)-m_{b}(t)\right\Vert =\min_{i}\left\Vert x_{j}(t)-m_{i}(t)\right\Vert .\label{eq:SOTM_match}
\end{equation}

Then each reference vector $m_{i}(t)$ is adjusted using the batch
update formula: 

\begin{equation}
m_{i}(t)=\frac{\sum_{j=1}^{N(t)}h_{ib(j)}(t)x_{j}(t)}{\sum_{j=1}^{N(t)}h_{ib(j)}(t)},\label{eq:SOTM_update}
\end{equation}

where index $j$ indicates the input data that belong to unit $b$
and the neighborhood function $h_{ib(j)}(t)\in\left(0,1\right]$ is
defined as a Gaussian function

\begin{equation}
h_{ib(j)}(t)=\exp\left(-\frac{\left\Vert r_{b}(t)-r_{i}(t)\right\Vert ^{2}}{2\sigma^{2}}\right),\label{eq:SOTM_gaussian}
\end{equation}

where $\left\Vert r_{b}(t)-r_{i}(t)\right\Vert ^{2}$ is the squared
Euclidean distance between the coordinates of the reference vectors
$m_{b}(t)$ and $m_{i}(t)$ on the one-dimensional array, and $\sigma$
is the user-specified neighborhood parameter. In contrast to what
is common for the standard batch \ac{SOM}, the neighborhood $\sigma$
is constant over time for a comparable timeline, not a decreasing
function of time as is common when time represents iterations.

\subsection*{}

\begin{acronym}[CRISP-DMD]
\acro{AE}{advanced economy}
\acro{AUC}{area under the curve}
\acro{AI}{artificial intelligence}
\acro{ANN}{artificial neural network}
\acro{AANN}{Auto-Associative Neural Network}
\acro{BIS}{Bank for International Settlements}
\acro{BMU}{best-matching unit}
\acro{BSC}{Banking Supervision Committee}
\acro{CCA}{Curvilinear Component Analysis}
\acro{CCE}{Coordinated Compilation Exercise}
\acro{CCF}{Correct Classification Frontier}
\acro{CDA}{Curvilinear Distance Analysis}
\acro{CDS}{credit default swap}
\acro{CHAPS}{Clearing House Automated Payment System}
\acro{CISS}{Composite Indicator of Systemic Stress}
\acro{CRISP-DM}{Cross Industry Standard Process for Data Mining}
\acro{DA}{Discriminant Analysis}
\acro{DDR}{data-dimension reduction}
\acro{DQRS}{Data Quality Reference Site}
\acro{EME}{emerging market economy}
\acro{ECB}{European Central Bank}
\acro{EU}{European Union}
\acro{XOM}{Exploration Observation Machine}
\acro{EDA}{exploratory data analysis}
\acro{FDI}{Financial Distress Index}
\acro{FSI}{financial soundness indicator}
\acro{FCM}{Fuzzy c-means}
\acro{GA}{Genetic Algorithm}
\acro{GDDS}{General Data Dissemination System}
\acro{GDP}{gross domestic product}
\acro{GFSM}{Global Financial Stability Map}
\acro{GTM}{Generative Topographic Mapping}
\acro{IA}{intelligence amplification}
\acro{IMF}{International Monetary Fund}
\acro{IS}{information systems}
\acro{IT}{information technology}
\acro{KD}{knowledge discovery}
\acro{KDD}{knowledge discovery in databases}
\acro{LE}{Laplacian Eigenmaps}
\acro{LMDS}{Local MDS}
\acro{LLE}{Locally Linear Embedding}
\acro{FSOM}{Feedback SOM}
\acro{MDS}{Multidimensional Scaling}
\acro{MPI}{macro-prudential indicator}
\acro{MSOM}{Merge SOM}
\acro{MVU}{Maximum Variance Unfolding}
\acro{NG}{neuro-genetic}
\acro{OECD}{Organisation for Economic Co-operation and Development}
\acro{PCA}{Principal Component Analysis}
\acro{QE}{quantization error}
\acro{ROC}{Receiver Operating Characteristics}
\acro{RSOM}{Recurrent SOM}
\acro{RecSOM}{Recursive SOM}
\acro{RO}{research objective}
\acro{RT}{research theme}
\acro{RQ}{research question}
\acro{SDDS}{Special Data Dissemination Standard}
\acro{SIFI}{systemically important financial institutions}
\acro{SNA}{System of National Accounts}
\acro{SOFSM}{Self-Organizing Financial Stability Map}
\acro{SOM}{Self-Organizing Map}
\acro{SOMSD}{SOM for structured data}
\acro{SOTM}{Self-Organizing Time Map}
\acro{t-SNE}{t-distributed Stochastic Neighbor Embedding}
\acro{TARGET}{Trans-European Automated Real-time Gross Settlement Express Transfer System}
\acro{TSOM}{Temporal SOM}
\acro{TPM}{transition probability matrix}
\acro{UK}{United Kingdom}
\acro{US}{United States}
\acro{VQ}{Vector Quantization}
\end{acronym} 
\end{document}